\documentclass[aip,reprint]{revtex4-1}
\DeclareUnicodeCharacter{0301}{\'e}
\usepackage{graphicx}% Include figure files
\usepackage{dcolumn}% Align table columns on decimal point
\usepackage{bm}% bold math
\usepackage[utf8]{inputenc}
\usepackage[T1]{fontenc}
\usepackage{mathptmx}
\usepackage{etoolbox}
\usepackage{xcolor}
\usepackage{placeins} 
\usepackage{amsmath} 
\usepackage{setspace} % For spacing control

\makeatletter
\def\@email#1#2{%
 \endgroup
 \patchcmd{\titleblock@produce}
  {\frontmatter@RRAPformat}
  {\frontmatter@RRAPformat{\produce@RRAP{*#1\href{mailto:#2}{#2}}}\frontmatter@RRAPformat}
  {}{}
}%
\makeatother
\begin{document}

\preprint{AIP/123-QED}

\title{Influence of Aspect Ratio and Flow Compressibility on Flow Dynamics in a Confined Cavity}
% Force line breaks with \\
\author{Sreejita Bhaduri}\affiliation{%
Department of Aerospace Engineering, Indian Institute of Technology Kanpur, 208016, Kanpur, India%%\\This line break forced% with \\
}
 %\altaffiliation[Also at ]{Department of Aerospace Engineering, Indian Institute of Technology Kanpur, India}%Lines break automatically or can be forced with \\

\author{Mohammed Ibrahim Sugarno}
 %\homepage{http://www.Second.institution.edu/~Charlie.Author.}
 \affiliation{%
Department of Aerospace Engineering, Indian Institute of Technology Kanpur, 208016, Kanpur, India%%\\This line break forced% with \\
}
\author{Ashoke De}%
 \email{ashoke@iitk.ac.in.}
 \affiliation{%
Department of Aerospace Engineering, Indian Institute of Technology Kanpur, 208016, Kanpur, India%%\\This line break forced% with \\
}
%\affiliation{ 
%Department of Sustainable Energy Engineering, Indian Institute of Technology Kanpur, 208016, %Kanpur, India%\\This line break forced with \textbackslash\textbackslash
%}

\date{\today}% It is always \today, today,

\begin{abstract}  

Cavities possess self-sustaining oscillations driven by the interaction of hydrodynamic and acoustic characteristics. These oscillations have applications in fuel-air mixing, heat exchangers, and landing gears, but resonance can damage the structures that house the cavities. Consequently, understanding cavity oscillations under varying geometries and flow conditions is essential for optimizing their benefits while minimizing the adverse effects. The present study investigates flow variations in a supersonic cavity confined by a top wall with a fixed deflection angle of $2.29^\circ$. We examine two aspect ratios of the cavity across freestream Mach numbers from 1.71 to 3 using Large-Eddy Simulations (LES) in OpenFOAM. Numerical Schlieren reveals the key flow structures, while spectral analysis and reduced-order modeling help identify dominant frequency modes and the corresponding flow structures. The results show that the shock from the deflection corner induces high gradients in the flow properties as it impinges on the shear layer. This amplifies the Kelvin-Helmholtz (KH) instability, which enhances mixing and mitigates the expected increase in oscillation frequency with Mach number. The KH instability develops spatially. Hence, the location of the shock impingement on the shear layer and the distance that the disturbances in the shear layer convect before reaching the cavity wall significantly influence the prominence of the instability, thereby influencing the frequency of cavity oscillations.
\end{abstract}

\maketitle

%\begin{quotation}

%\end{quotation}
\begin{figure*}

	\includegraphics[scale=0.48]{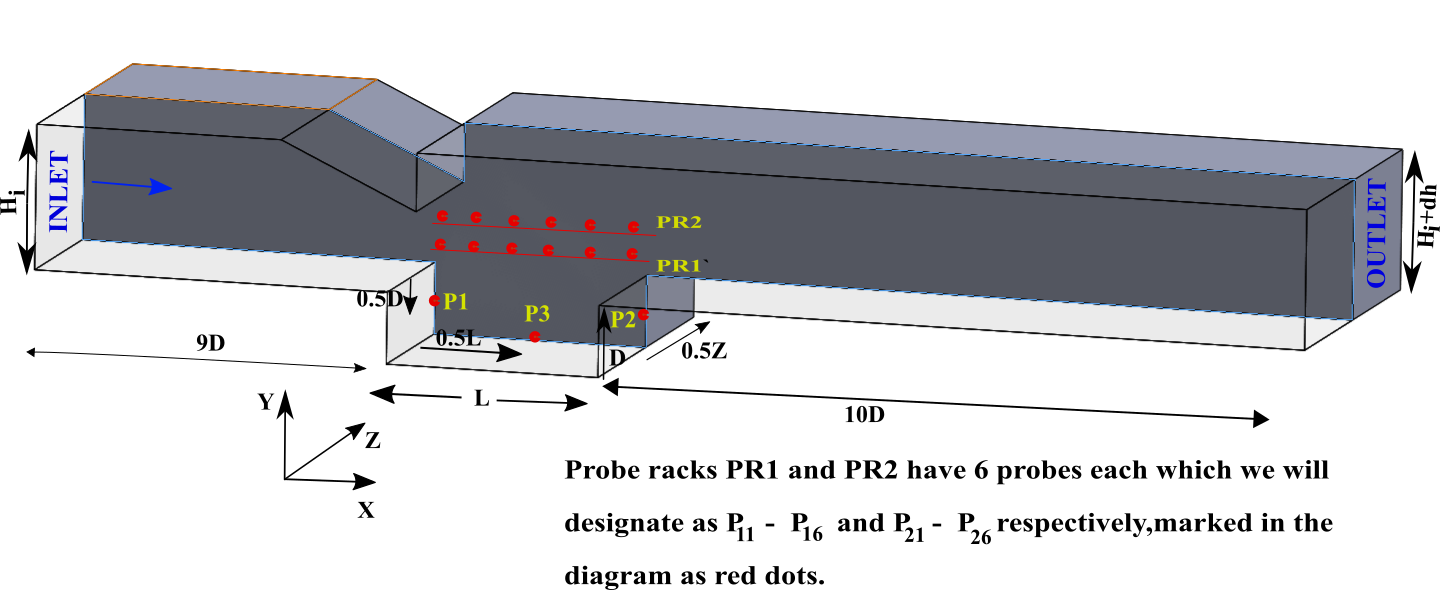}% Here is how to import EPS art
    \centering

    \caption{\label{fig:1} Schematic of the cavity with probe locations. The probes, P1, P2, and P3, are located at the midpoints of the front wall, aft wall, and floor of the cavity, as indicated by the red dots. PR1 and PR2 denote probe racks at y = 0.1D and y = 0.5D, respectively. Each rack has 6 probes, labeled individually as P$_{Nn}$, where N indicates the rack number and n represents the probe index, with 1 located nearest to the leading edge and 6 closest to the trailing edge. Table \ref{tab:table1} provides detailed probe locations.}
   \end{figure*}
   \begin{figure}

	\includegraphics[scale=0.27]{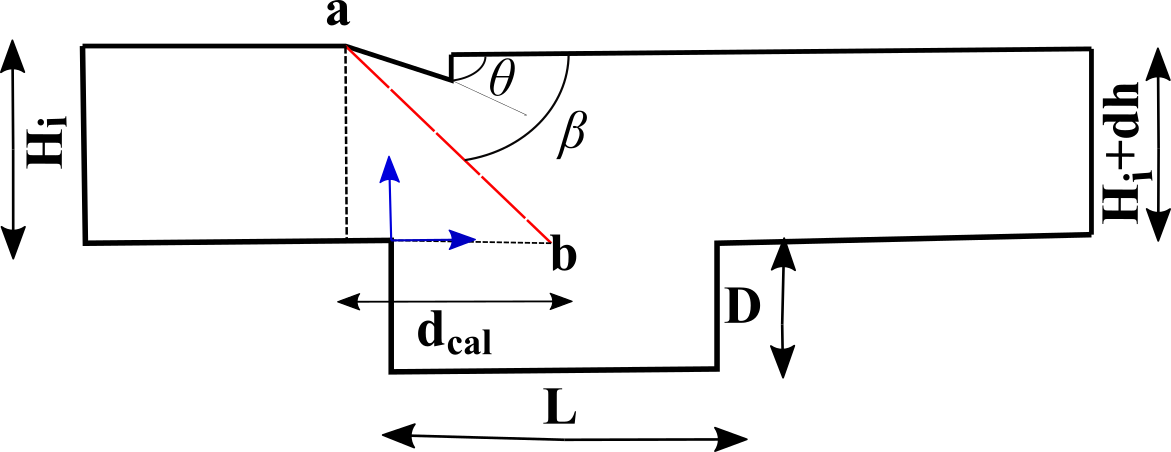}% Here is how to import EPS art
    \centering

    \caption{\label{fig:2} Schematic of cavity geometry illustrating the deflection angle ($\theta$), the shock angle ($\beta$), the origin of the shock (a), and the impinging point (b).}
   \end{figure}
    \begin{figure}

	\includegraphics[scale=0.425]{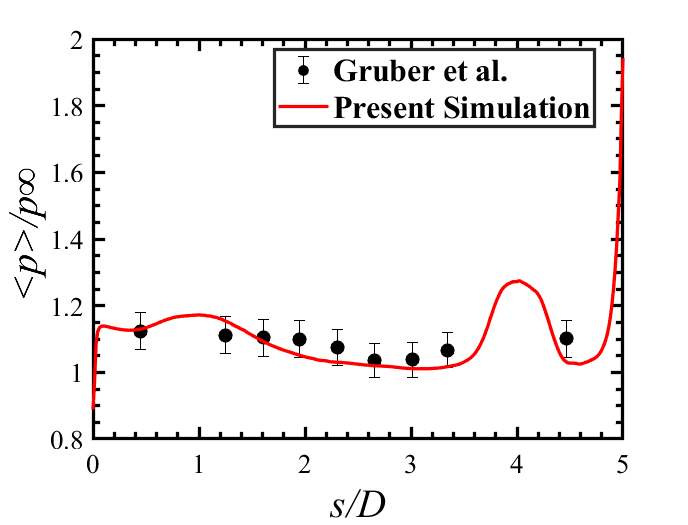}% Here is how to import EPS art
    \centering

    \caption{\label{fig:3} Validation of the present simulation against the experimental data of Gruber et al.\cite{gruber2001fundamental}.}
   \end{figure}

\section{\label{sec:level1}Introduction}
Cavities' self-sustaining oscillations make them useful as they improve fluid mixing by enabling momentum and energy exchange. They are also an interesting research topic because of their intricate flow physics\cite{gruber2001fundamental,sitaraman2021adaptive,johnson2010instability,chakravarthy2018analytical,devaraj2020experimental,sekar2020unsteady}. They play a significant role in air-fuel mixing in scramjet combustors and thermal management in spacecraft and aircraft. They are widely employed in industrial systems such as weapon bays, fuel tanks, resonator nozzles, and heat-exchangers\cite{charwat1961investigation,emery1969recompression,saravanan2020isolator}. Additionally, cavities serve as thermal barriers in insulation, promote coolant flow in nuclear reactors, and improve heat transfer in electronics cooling and solar energy systems\cite{swift2017thermoacoustics,choi1995enhancing,zalba2003review}.

Rossiter\cite{rossiter1964wind} characterized cavity oscillations as a feedback-driven phenomenon wherein instabilities in the shear layer grow spatially, impinge on the trailing edge, and induce pressure waves that propagate upstream, reinforcing shear layer perturbations by an acoustic-vortex resonance mechanism. Numerous studies have validated Rossiter’s definition, expanding it with insights gained through advancements in technology and physics \cite{heller1971flow,heller1975physical,heller1996cavity,krishnamurty1955acoustic}. Rowley\cite{rowley2006dynamics} also identified two primary cavity oscillation modes: the shear layer mode, where vortices roll up in the shear layer and impinge on the aft edge, and the wake mode, prevalent in incompressible flows, characterized by large-scale shedding from the leading edge toward the cavity. Geometrical parameters like the L/D ratio and extrinsic factors like the thickness, type, and freestream Mach number of the boundary layer strongly influence cavity dynamics\cite{plentovich1993experimental,tracy1992measurements}. Stallings \cite{stallings1987experimental} classified supersonic cavities into open cavities (L/D < 10) and closed cavities (L/D > 13), each exhibiting distinct flow features\cite{wang2013characteristics,heller1971flow}.Open cavities, prevalent in aerospace applications, rely on a balance between freestream energy input and energy dissipation through acoustic radiation, viscous losses, and convective mass exchange. Conversely, closed cavities, with their higher drag coefficients, are more suited for heat exchange applications. 

Lawson and Barakos \cite{lawson2011review} pointed out that the closed cavities lack significant acoustic features. Conversely, open cavities have two main components that define their acoustic signature: a) discrete tones (Rossiter modes) generated by interactions like shear-wall, shock-shear layer, vortex-vortex, vortex-wall, or vortex-shear layer interactions, and b) low-energy broadband noise from turbulent fluctuations, shear layers, and freestream interactions. The oscillations present in the open cavities hence contribute to the acoustic loading on the structures housing them, potentially leading to structural damage \cite{rockwell1978self,rowley2006dynamics}. Although these oscillations are critical for cavity applications, it is essential to develop control techniques to regulate their frequencies. Previous investigations have explored flow control methods for open supersonic cavities without confinement \cite{alam2007new,malhotra2016aft,lad2018experimental,jain2021aero,jain2023effects,lee2008passive}. However, in practical applications, such as those in scramjet combustors, the cavities are confined by a top wall. In these confined cavities, multiple shock reflections occur between the confinement walls and the shear layer. The incident shock impinges on the separating shear layer as it convects disturbances downstream, altering the flow field within and above the cavity. Therefore, analyzing the impact of shock impingement on the shear layer is essential for a detailed study of cavity oscillations and the development of control mechanisms.

Karthick \cite{karthick2021shock} investigated the effects of varying shock strengths on a confined supersonic cavity (L/D = 2, Mach 1.71) using two-dimensional Detached Eddy Simulations (DES). His findings revealed that shock-shear layer interactions altered wall pressure distributions and oscillation frequencies. He introduced a confined supersonic wall jet for flow control, reducing jet-column instability by 58\% and improving flow stability. In one of our recent studies\cite{bhaduri2024flow}, we explored passive control of cavity oscillations in a two-dimensional cavity (L/D = 3) confined by a top wall with a $3.6^\circ$ deflection angle across a flow of Mach 1.71. The URANS simulations demonstrated significant suppression of oscillation by the sub-cavities incorporated at the front and aft walls of the cavity. However, these studies have not fully accounted for Mach number variations or three-dimensional effects.

A comprehensive understanding of cavity dynamics necessitates three-dimensional simulations, which better capture turbulence mechanisms such as vorticity stretching and intricate flow interactions. Furthermore, the Mach number of the flow significantly impacts cavity oscillations, necessitating a more detailed analysis of its influence\cite{cant2001sb,george2013lectures}. To address these gaps, we conducted Large Eddy Simulations (LES) of supersonic open cavities with and without top-wall confinement\cite{bhaduri2024effects}. Our results show that the impinging shock from the top wall induces Kelvin-Helmholtz instability, enhancing mixing while reducing oscillation frequency. A preliminary analysis of Mach number variation of $\delta M = 0.3 $, indicated that a higher Mach number increases oscillation frequency due to the higher speed of the flow. However, this increase is less pronounced in the presence of an impinging shock.

The present investigation builds upon our previous investigations by systematically varying M$_\infty$ over a broader range while maintaining a fixed deflection angle of the top wall. This adjustment alters the angle of the impinging shock, thereby shifting the impingement location on the shear layer. Since the aspect ratio (L/D) of the cavity greatly influences cavity flow, we also examine how it affects flow characteristics, which was not considered in our earlier investigation.  This study specifically aims to:

\begin{itemize}
\item {Investigate the effect of Mach number (1.71 $\leq M_\infty \leq$ 3.0) on flow characteristics and oscillation frequency in a confined open cavity with an impinging shock from a $2.29^\circ$ deflection corner.}
\item {Analyze how the cavity aspect ratio influences flow physics and oscillation frequency across the considered Mach number range. We are varying the aspect ratio by changing the length of the cavity.}
\end{itemize}

The article is structured as follows: Section \ref{sec:level2} covers the geometrical configuration, numerical methods, boundary conditions, grid independence study, and validation against experiments. Section \ref{sec:level3} details the results and their physical interpretation, discussing the variation in flow field and frequency content with Mach numbers and aspect ratios. Section \ref{sec:level4} summarizes the key findings of this numerical investigation.

\section{\label{sec:level2}Numerical Methodology}

\subsection{Geometrical Configuration}
Figure \ref{fig:1} presents a schematic of the cavity, positioned 9D from the inlet to ensure a fully developed incoming turbulent flow. The outlet extends 10D downstream of the trailing edge to facilitate smooth flow through the domain. The inlet height is 2.021D, while the outlet height is 1.954D, which accounts for the deflection of the top wall that generates the impinging shock. The computational domain extends 5D in the spanwise direction ($Z=5D$) to capture three-dimensional effects.
The probes are placed in the midplane of the spanwise direction ($0.5Z=2.5D$) to analyze the evolving flow conditions. They are placed on the walls of the cavity and within the computational domain to capture key flow events. Inside the cavity, P1, P2, and P3 are located at the front wall, aft wall, and floor midpoints, respectively, as shown by the red dots in figure \ref{fig:1}. Two probe racks, PR1 at $y=0.1D$ and PR2 at $y=0.5D$, track shear layer oscillations under different flow and geometric conditions. Each rack contains six probes, labeled sequentially, as $P_{N1}$ to $P_{N6}$ in the flow direction, where $N=1$ for PR1 and $N=2$ for PR2. Table \ref{tab:table1} elucidates the coordinates of all the probe points used in the present work.
% Table 1
\begin{table}[ht]
    \centering
    \caption{\label{tab:table1} Coordinates of the internal probes and probes in probe rack PR1 ($y=0.1D$) ($N=1$) and probe rack PR2 ($y=0.5D$) ($N=2$). All the coordinates are expressed in terms of the length (L) and depth (D) of the cavity.}
    \begin{ruledtabular}
    \begin{tabular}{cc}
        \multicolumn{2}{c}{\textbf{Internal Probes}} \\
        \hline
        \textbf{Probe} & \textbf{Coordinates} \\
         \hline
      \mbox{P1} & \mbox{(0, -0.5D, 2.5D)} \\  % Blank row for user to fill
      \mbox{P2} & \mbox{(L, -0.5D, 2.5D)} \\  % Blank row for user to fill
       \mbox{P3} & \mbox{(-0.5L, -0.5D, 2.5D)} \\ % Blank row for user to fill
       \hline
    \multicolumn{2}{c}{\textbf{Probe Racks}} \\  
        \hline
        \textbf{Probe number ($P_{Nn}$)} & \textbf{Coordinates} \\  
        \hline
        \mbox{n=1} & \mbox{(-D, y, 2.5D)} \\
        \mbox{n=2} & \mbox{(-0.5D, y, 2.5D)} \\
        \mbox{n=3} & \mbox{(0, y, 2.5D)} \\
        \mbox{n=4} & \mbox{(0.5D, y, 2.5D)} \\
        \mbox{n=5} & \mbox{(D, y, 2.5D)} \\
        \mbox{n=6} & \mbox{(2D, y, 2.5D)} \\
    \end{tabular}
    \end{ruledtabular}
\end{table}

Figure \ref{fig:2} illustrates how the impingement point shifts as the shock angle ($\beta$) varies with the freestream Mach number (M$_\infty$) while maintaining a constant deflection angle ($\theta$). The schematic in the x-y plane provides a clearer visualization of these parameters. According to the $\theta$-$\beta$-M$_\infty$ relation, increasing M$_\infty$ with a fixed $\theta$ decreases $\beta$. In this analysis, we set the deflection angle to $2.29^\circ$ and observe the reduction in $\beta$ as M$_\infty$ increases.

The origin (0,0) in Figure \ref{fig:2} lies at the leading edge of the cavity. To examine the origin of the impinging shock, its impingement point on the shear layer, and its variation with flow conditions, we define the shock’s origin at the deflection corner on the top wall as ‘a’ and the impingement point on the shear layer as ‘b,’ as marked in Figure \ref{fig:2}. The red line represents the impinging shock. Point 'b' provides a preliminary estimate of the approximate impingement distance from the leading edge. As M$_\infty$ increases, point 'b' moves downstream.

Table \ref{tab:table2} presents $\beta$ with the corresponding M$_\infty$, the calculated distance $d_{cal}$ of point 'b' from the origin, expressed in terms of the cavity depth (D). For the cavity with $L/D = 3$ at M$_\infty $= 3, $d_{cal}$ reaches 3.71D, extending beyond the trailing edge. Similarly, for $L/D = 2$, the impingement points at M$_\infty$ = 2.29 and 3 also extend beyond the trailing edge. Given the supersonic nature of the flow, the shock’s impingement on the shear layer in these cases is unlikely to significantly alter the cavity flow dynamics. However, the interaction of the impinging shock with the separation shock at the leading edge shifts the impingement point upstream, as seen in Table \ref{tab:table2}. The calculation  of $d_{cal}$ does not account for this separation shock’s influence.

In table \ref{tab:table2},  $d_{obt}$ represents the streamwise distance of the impingement point on the shear layer at the instant marking the beginning of a feedback loop, as obtained from the simulations. This streamwise distance remains nearly constant throughout the entire cycle. Section \ref{section:flow} provides further details in the flow visualization analysis. Therefore, except for the $L/D = 2$ cavity at M$_\infty = 3$, the impinging shock significantly influences the cavity dynamics. In the following sections, we examine this upstream shift in the impingement location as a function of M$_\infty$ and the cavity aspect ratio in greater detail.

\begin{table}[ht]
    \centering
    \caption{\label{tab:table2} Varying distance of the impinging point from the origin, calculated ($d_{cal}$) and obtained ($d_{obt}$) from simulations, with varying \( M_\infty \) and \( \beta \) for L/D = 3 and L/D = 2.}
    
    \begin{ruledtabular}
        \begin{tabular}{ccccc}
            \textbf{\( M_\infty \)} & \textbf{\( \beta \)} & \textbf{$d_{cal}$} & \textbf{$d_{obt}$\newline ( \( L/D = 3 \))} & \textbf{$d_{obt}$\newline ( \( L/D = 2 \))} \\
            \hline
            1.71 & 37.95  & 1.06D  & 0.406D & 0.59D  \\
            2.00 & 31.89  & 1.72D  & 0.84D  & 1.17D  \\
            2.29 & 27.65  & 2.33D  & 1.14D  & 1.742D \\
            3.00 & 21.08  & 3.71D  & 2.78D  & 2.87D  \\
        \end{tabular}
    \end{ruledtabular}
\end{table}

\subsection{Governing equations}

The simulations use the Favre-averaged (density-weighted) filtered Navier-Stokes equations (Eqs.\ref{eq:continuity}, \ref{eq:momentum},and \ref{eq:energy}) within the framework of Large Eddy Simulation (LES) to analyze the compressible flow. LES is a method widely employed to model turbulent flow fields. It resolves large-scale eddies directly while modeling small-scale eddies through various sub-grid scale (SGS) models. Filtering primarily decomposes the velocity field into resolved and subgrid-scale components \cite{moin1982numerical,piomelli1999large,cant2001sb,george2013lectures}.
The filtered equations are analogous to the standard Navier-Stokes equations but include the SGS stress tensor, which is modeled using an eddy viscosity approach to close the equations. The working medium is assumed to adhere to the ideal gas laws. The governing equations employed in the present simulations are listed below \cite{soni2019modal,arya2021effect}:
\begin{equation}
\frac{\partial \overline{\rho}}{\partial t} +
\frac{\partial}{\partial x_i}\left[ \overline{\rho} \widetilde{u_i} \right] = 0
\label{eq:continuity}
\end{equation}

%\begin{equation}
%\frac{\partial}{\partial t}\left( \overline{\rho} \widetilde{u_i} \right) +
%\frac{\partial}{\partial x_j}
%\left[
%\overline{\rho} \widetilde{u_j} \widetilde{u_i}
%+ \overline{p} \delta_{ij}
%- \widetilde{\tau_{ji}^{tot}}
%\right]
%= 0
%\label{eq:momentum}
%\end{equation}
\begin{equation}
\frac{\partial}{\partial t} \left( \overline{\rho} \widetilde{u_i} \right) +
\frac{\partial}{\partial x_j} \left( \overline{\rho} \widetilde{u_i} \widetilde{u_j} \right) =
- \frac{\partial}{\partial x_i} \left( \overline{p} \right) +
\frac{\partial}{\partial x_j} \left[ (\mu + \mu_t) \frac{\partial \widetilde{u_i}}{\partial x_j} \right]   \label{eq:momentum}
\end{equation}

\begin{align}
  \frac{\partial}{\partial t} 
  \left( \overline{\rho} \widetilde{E} \right) +
  \frac{\partial}{\partial x_i} 
  \left( \overline{\rho} \widetilde{u_i} \widetilde{E} \right) &= 
  - \frac{\partial}{\partial x_j} 
  \left[ \widetilde{u_j} \left( - \widetilde{p} I + (\mu + \mu_t) 
  \frac{\partial \widetilde{u_i}}{\partial x_j} \right) \right] \notag \\ 
  &\quad + \frac{\partial}{\partial x_i} 
  \left[ \left( k + \mu_t C_p \text{Pr}_t \right) \frac{\partial \widetilde{T}}{\partial x_i} \right]
  \label{eq:energy}
\end{align}

where \( \overline{(.)} \) denotes time-averaged parameters, and \( \widetilde{(.)} \) represents density-averaged parameters.$\rho$ is the density, $u_i$ is the velocity vector, $p$ is the pressure and 
$E = e + \frac{u_i^2}{2}$ is the total energy of the system, where $e = \frac{h - p}{\rho}$ 
is the internal energy; and $h$ is the enthalpy.  
$\mu$ and $k$ represent the viscosity and thermal conductivity of the fluid, respectively, 
and $\mu_t$ and $\text{Pr}_t$ denote the  eddy viscosity and 
the turbulent Prandtl number, respectively.
  
 Wall Adapting Local Eddy (WALE) viscosity model\cite{nicoud1999subgrid}  determines the eddy viscosity ($\mu_t$). This model is based on the resolved velocity field's strain and rotation rate (Eq. \ref{eq:mu}) and offers improved predictions of the turbulent flows, especially close to the walls. The equation evaluating the eddy viscosity is as follows:

\begin{equation} \label{eq:mu}
\mu_{t} = \rho \left( C_w \cdot \Delta_s \right)^2 
\frac{\left( S_{ij}^{d} S_{ij}^{d} \right)^{3/2}}
{\left( \widetilde{S}_{ij} \widetilde{S}_{ij} \right)^{5/2} + \left( S_{ij}^{d} S_{ij}^{d} \right)^{5/4}}
\end{equation}

 where $\widetilde{S}_{ij} = \frac{1}{2} (\widetilde{g}_{ij} + \widetilde{g}_{ji} ), \quad
    \widetilde{g}_{ij} = \frac{\partial \widetilde{u}_i}{\partial x_j}$ and ($\Delta_s$) is the filter width. \(C_w\) is a pure constant, and in the present simulation it is assumed to be 0.5.

\subsection{Numerical schemes and Validation.}
A finite volume, density-based solver, rhoCentralRK4Foam \cite{li2020scalability}, solves the governing equations within the OpenFOAM framework \cite{jasak2009openfoam}. It uses a third-order accurate, four-stage, low-storage Runge-Kutta method for explicit time integration. Convective fluxes are discretized using the central-upwind techniques of Kurganov and Tadmor \cite{kurganov2000new, adityanarayan2023leading}, while dissipative fluxes are handled with a second-order accurate central scheme. We use Sutherland's law to calculate viscosity. The polynomial formula from the Joint Army-Navy-Air Force (JANAF) model computes the specific heat of air \cite{chase1998nist}.
Readers are encouraged to refer to our recent publication \cite{bhaduri2024effects} for a detailed discussion on the grid independence study and LES validation, as these aspects are not included in this study to avoid redundancy.The findings indicate that the medium grid is computationally economical while satisfying LES resolution requirements. Experimental data from Gruber et al.\ \cite{gruber2001fundamental} is used to validate the flow solver. Their study investigated a supersonic flow of \( M_\infty = 1.71 \) over an open cavity with \( L/D \) ratio of 3, at a stagnation temperature of 300 K and a stagnation pressure of 690 kPa.  Time-averaged pressure distributions along the cavity walls (Figure \ref{fig:3}) further demonstrate that results obtained from the medium grid in the present simulation are within ±5\% tolerance compared to Gruber's experimental data.

For the ongoing study, the flow is initialized with a static pressure of 101325 Pa and a temperature of 189.29 K. We have maintained a constant pressure and temperature in all the flow and geometrical configurations throughout the study. The freestream velocities are varied to vary the freestream Mach numbers: M$_\infty$ = 1.71 (471 m/s), M$_\infty$ = 2 (551.576 m/s), M$_\infty$ = 2.29 (631.554 m/s), M$_\infty$ = 3 (828.683 m/s), and M$_\infty$ = 4 (1104.9 m/s). We introduce the initial fluctuations in the velocity field using the "Klein" inflow generator \cite{klein2003digital}. At the outlet, we extrapolate all variables under the assumption of supersonic flow, and we apply a "no-slip" condition on the insulated walls.

\section{\label{sec:level3}Results and Discussion}
\begin{figure*}

	\includegraphics[scale=1.4]{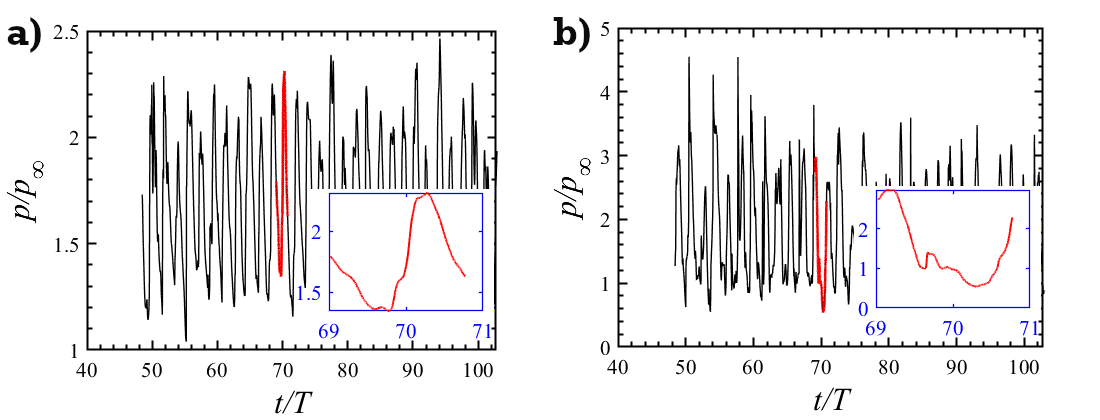}% Here is how to import EPS art
    \centering

    \caption{\label{fig:4} Temporal variation of pressure normalized with the freestream pressure (p/p$_\infty$) at the a) front and b) aft walls of the cavity of L/D = 3 at M$_\infty$ = 1.71. The time is normalized with T (L/U$_\infty$ = 8.26e-5 s).}
   \end{figure*}
   
\begin{figure*}
\includegraphics[scale=0.35]{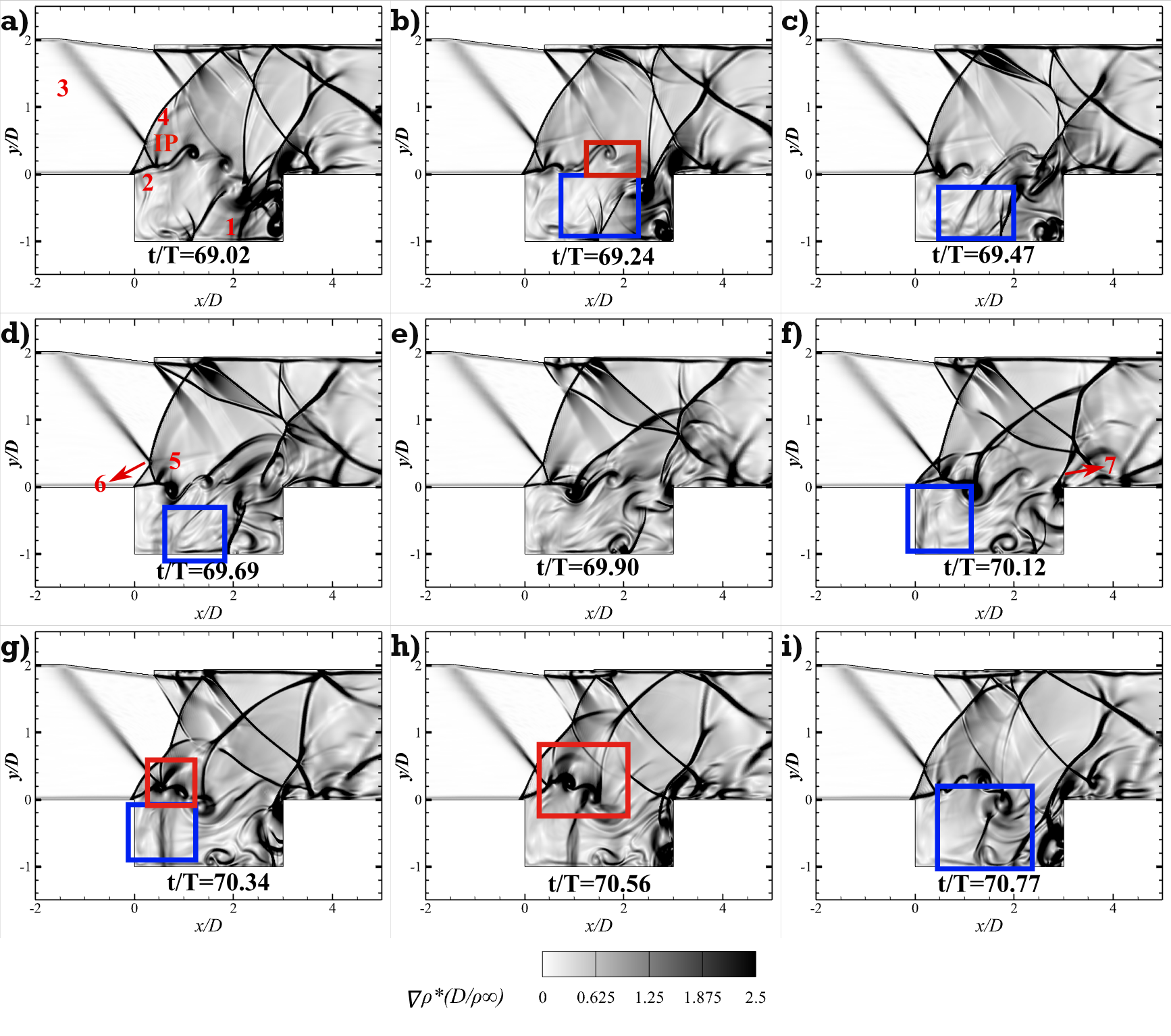}% Here is how to import EPS art
    \centering

    \caption{\label{fig:5} Normalized Density Gradient ($\nabla \rho \cdot (D/\rho_{\infty})$) contour from t/T=69.02 (a) to 70.77 (i) at intervals of 0.219 for cavity L/D=3 at M$_\infty$ = 1.71. Key flow features: (1) upstream traveling wave, (2) perturbed separating shear layer, (3) impinging shock, (4) separating shock wave, (5) expansion wave,  (6) interaction points of (3) and (4), and (7) reattachment shock. IP denotes the impinging point. The blue square highlights the upstream traveling pressure wave , and the red square marks the KH rolls.}
   \end{figure*}
     
   \begin{figure*}

	\includegraphics[scale=1.4]{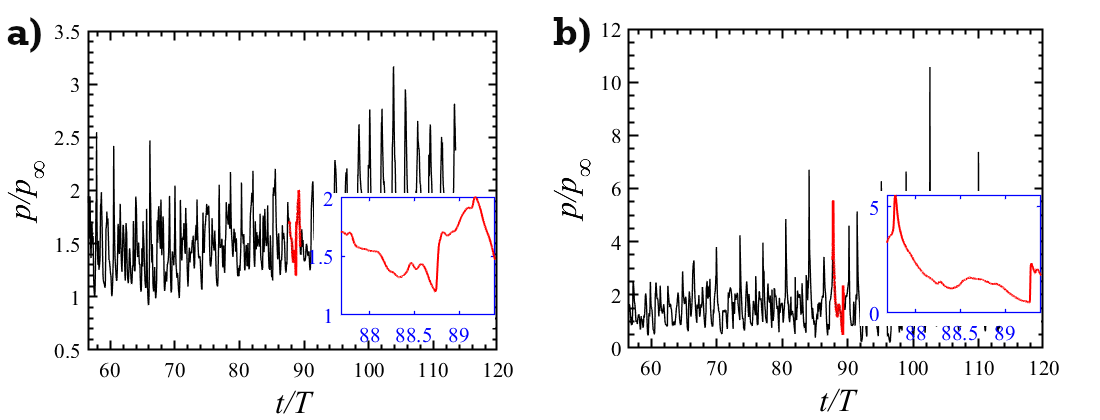}% Here is how to import EPS art
    \centering

    \caption{\label{fig:6} Temporal variation of pressure normalized with the freestream pressure (p/p$_\infty$) at the a) front and b) aft walls of the cavity of L/D = 3 at M$_\infty$ = 2. The time is normalized with T (L/U$_\infty$ = 7.071e-5 s).}
   \end{figure*}
   \begin{figure*}

	\includegraphics[scale=0.32]{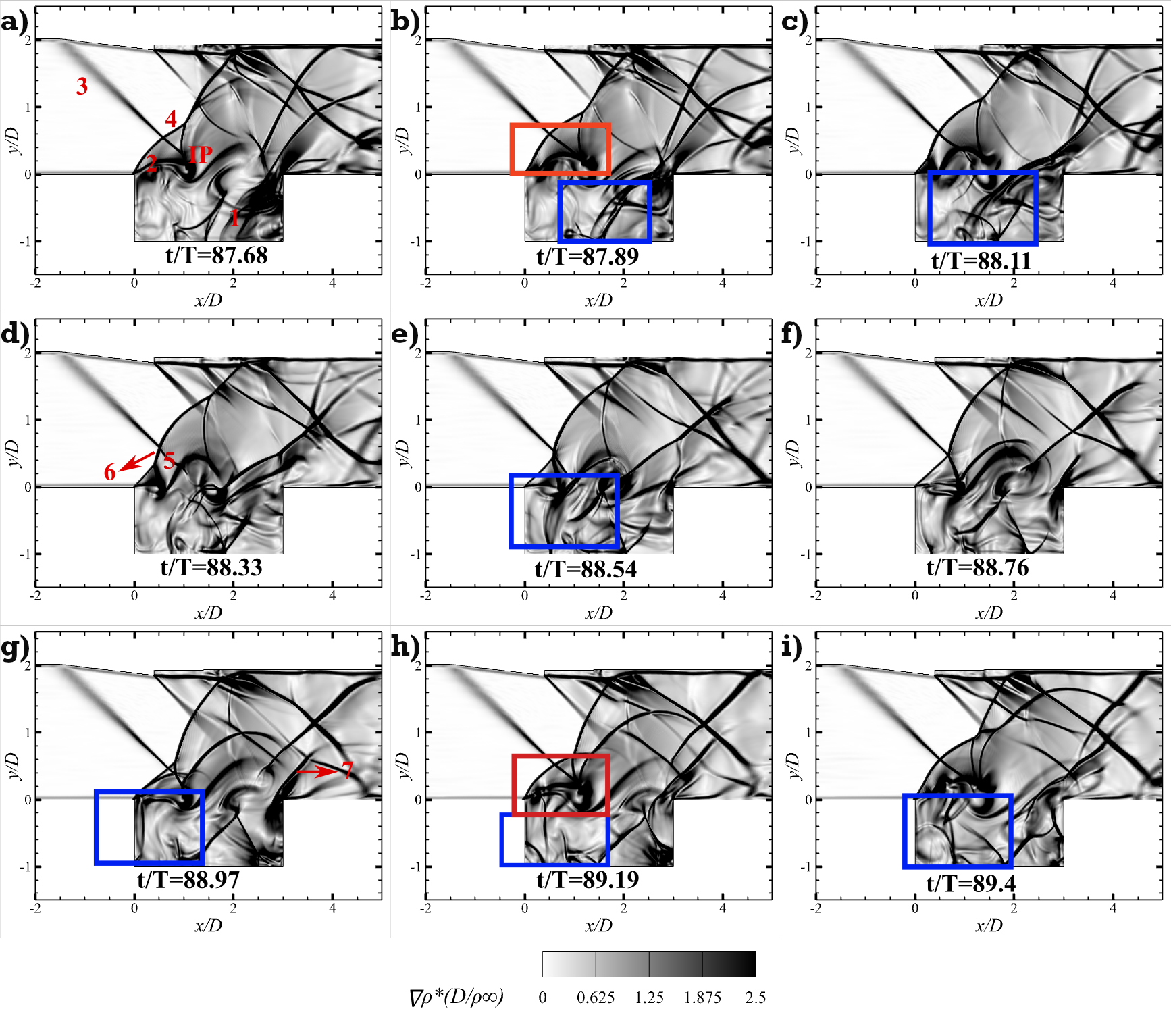}% Here is how to import EPS art
    \centering

    \caption{\label{fig:7} Normalized Density Gradient contour ($\nabla \rho \cdot (D/\rho_{\infty})$)  from the time step (t/T) of 87.68 (a) to 89.4(i) at an interval of 0.215 for the cavity of L/D = 3 at M$_\infty$ = 2. Key flow features: (1) the upstream traveling wave, (2) the perturbed separating shear layer, (3) the impinging shock, (4) the separating shock wave, (5) the expansion wave, (6) the interaction points of (3) and (4), and (7) reattachment shock. IP is the impinging point. The blue square marks the upstream traveling pressure wave. The red square demonstrates the KH rolls.}
   \end{figure*}
   \begin{figure*}

	\includegraphics[scale=1.4]{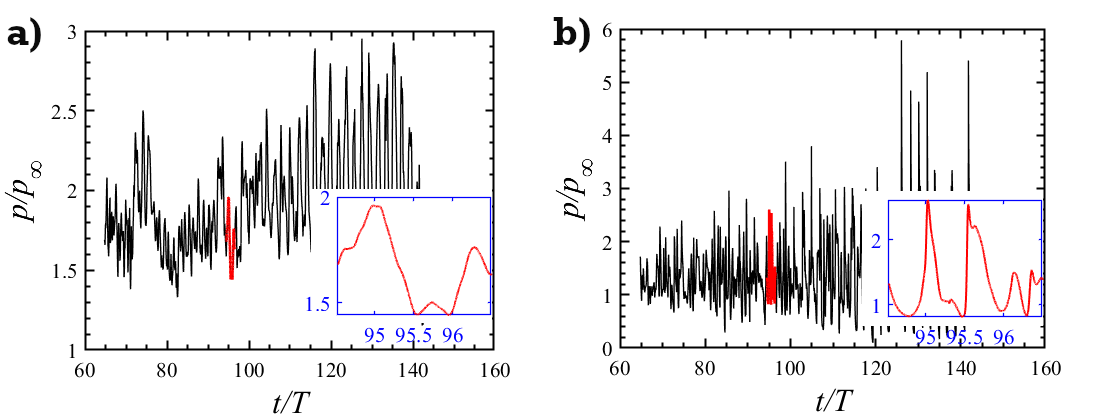}% Here is how to import EPS art
    \centering

    \caption{\label{fig:8} Temporal variation of pressure normalized with the freestream pressure (p/p$_\infty$) at the a) front and b) aft walls of the cavity of L/D=3 at M$_\infty$ = 2.29. The time is normalized with T (L/U$_\infty$ = 6.175e-5 s).}
   \end{figure*}
   \begin{figure*}

	\includegraphics[scale=0.32]{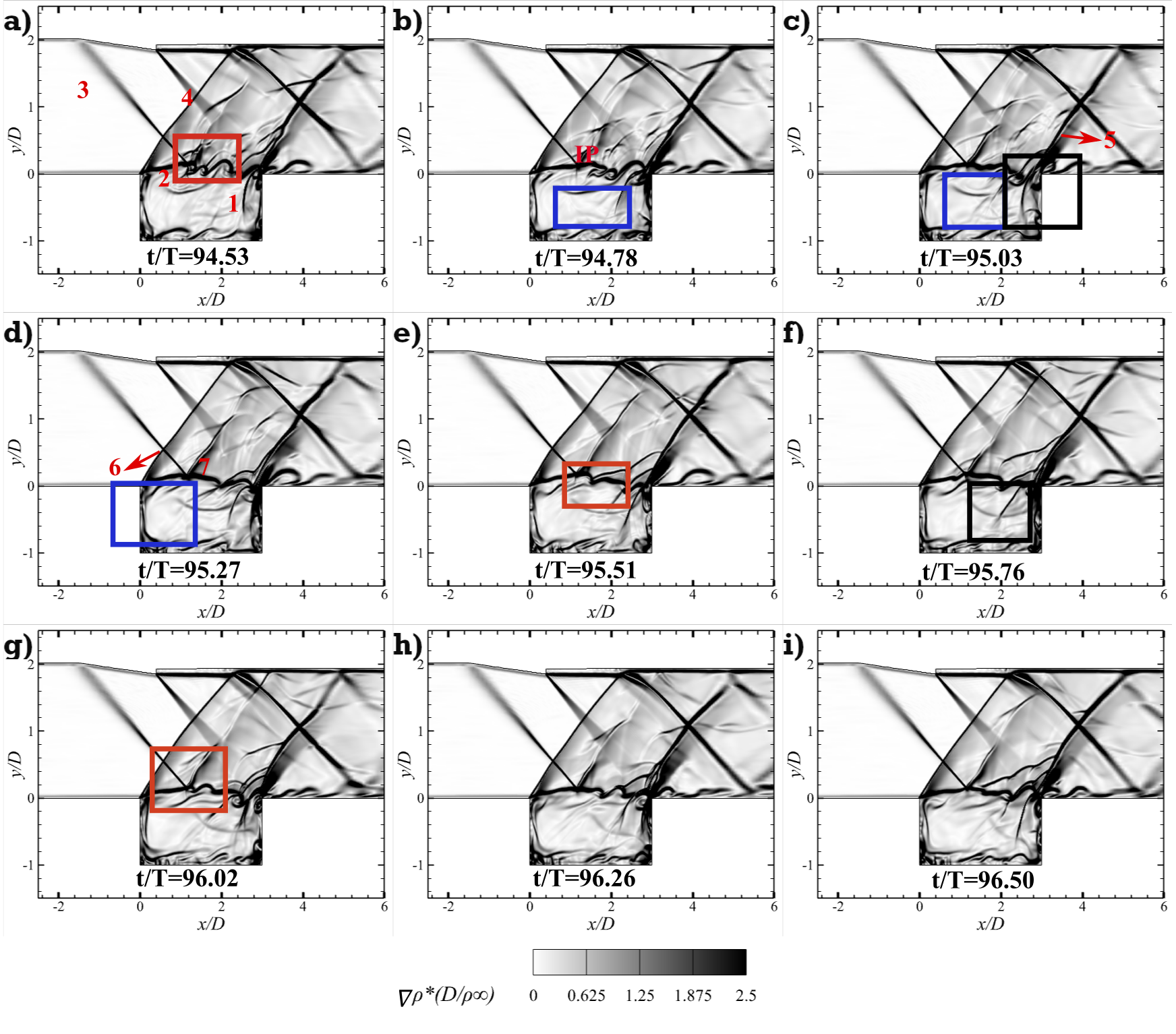}% Here is how to import EPS art
    \centering

    \caption{\label{fig:9} Normalized Density Gradient contour ($\nabla \rho \cdot (D/\rho_{\infty})$) from the time step (t/T) of 94.53 (a) to 96.5(i) at an interval of 0.246 for the cavity of L/D = 3 at M$_\infty$ = 2.29. Key flow features: (1) the upstream traveling wave, (2) the perturbed separating shear layer, (3) the impinging shock, (4) the separating shock wave, (5) reattachment shock,  (6) the interaction points of (3) and (4), and (7) the expansion wave. IP is the impinging point. The blue square marks the upstream traveling pressure wave. The red square demonstrates the KH rolls.}
   \end{figure*}
   
   \begin{figure*}

	\includegraphics[scale=1.4]{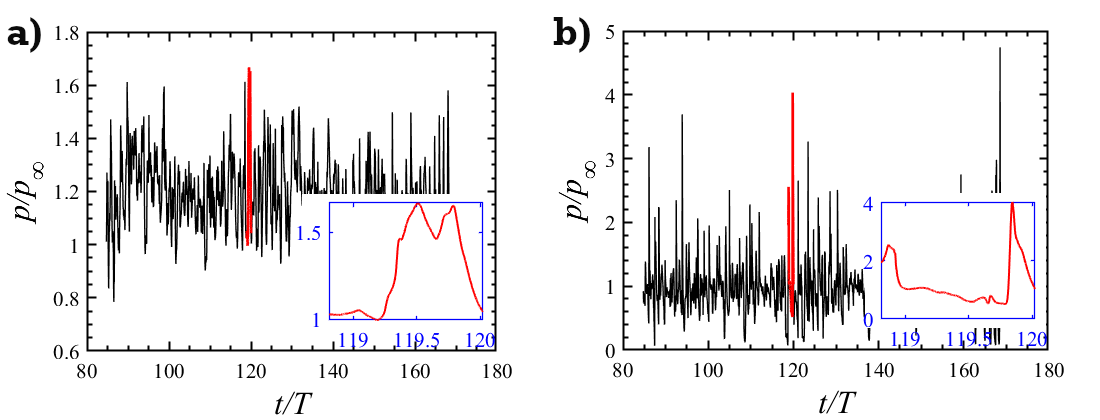}% Here is how to import EPS art
    \centering

    \caption{\label{fig:10} Temporal variation of pressure normalized with the freestream pressure (p/p$_\infty$) at the a) front and b) aft walls of the cavity of L/D=3 at M$_\infty$ = 3. The time is normalized with T (L/U$_\infty$ = 4.706e-5 s).}
   \end{figure*}
    \begin{figure*}

	\includegraphics[scale=0.32]{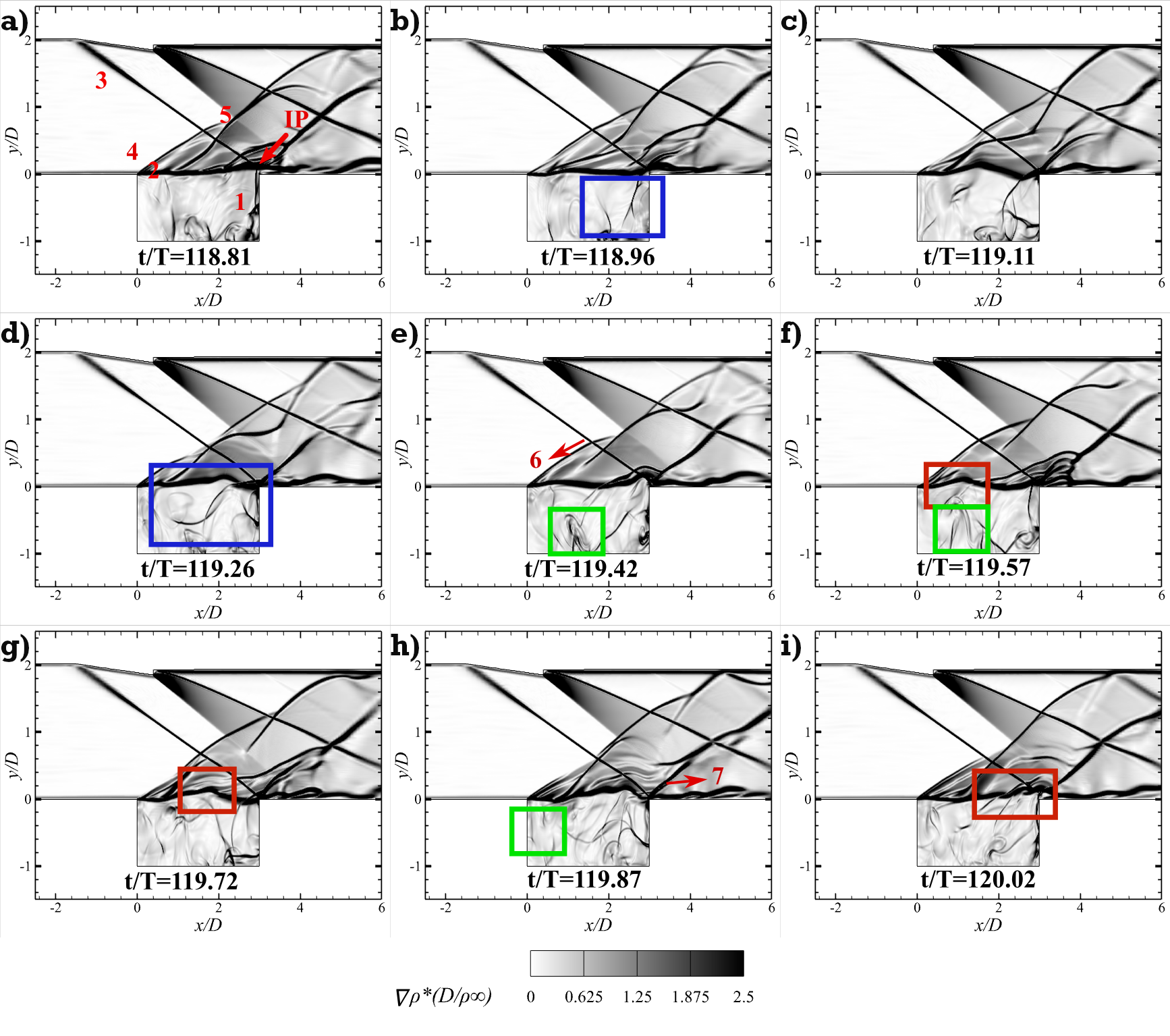}% Here is how to import EPS art
    \centering

    \caption{\label{fig:11} Normalized Density Gradient contour ($\nabla \rho \cdot (D/\rho_{\infty})$) from the time step (t/T) of 118.81 (a) to 120.02(i) at an interval of 0.151 for the cavity of L/D=3 at M$_\infty$=3. Key flow features: (1) the upstream traveling wave, (2) the perturbed separating shear layer, (3) the impinging shock, (4) the separating shock wave, (5) a disturbance wave,  (6) the interaction points of (3) and (4), and (7) reattachment shock. IP is the impinging point. The blue square marks the upstream traveling pressure wave. The red square demonstrates the disturbances in the shear layer, and the green square represents disturbances originating at the base of the cavity.}
   \end{figure*}
   % \begin{figure*}
   
This section has four subsections, each addressing different aspects of cavity flow physics for the two specified geometrical configurations under varying flow conditions. Numerical Schlieren visualization (Section \ref{section:flow}) provides insights into the flow structures across all the M$_\infty$ for both cavity configurations. Spectral analysis (Section \ref{section:SA}), including power spectral density and wavelet analysis, quantifies the oscillation frequencies and their dependence on flow and geometry. A comparison with Rossiter’s analytical solution validates the simulation results. Reduced-order modeling (Section \ref{section:DMD}) establishes the connection between spectral frequencies and flow structures.

\subsection{Flow Visualisation} \label{section:flow}
 In open cavity ($L/D \le 10$) flows, the shear layer separating from the leading edge convects disturbances as KH instability waves at specific frequency modes. These waves arise from the destabilizing effect of shear, which overcomes the stabilizing influence of density stratification. As the disturbances travel towards the trailing edge, they grow spatially and evolve into a periodic array of compact spanwise vortices. Through nonlinear interactions, these vortices merge and redistribute the vorticity field. As the disturbances impinge on the trailing edge, a reattachment shock forms and fluid enters the cavity. The entrainment of the mass inside the cavity results in a pressure wave that originates near the trailing edge inside the cavity and travels upstream. It induces the growth of the organized vortical structures in the shear layer via acoustic-vortex resonance. As these vortices merge or pair while convecting downstream, forming KH rolls, they enhance fluid entrainment, increasing the shear layer's spreading rate and aiding in momentum and heat transport \cite{heller1996cavity,krishnamurty1955acoustic,karthick2021shock,wang2013characteristics,thangamani2019mode,rockwell1978self}. Consequently, the feedback loop becomes integral to the cavity's applications. In the case of confined cavities, the shock wave from the top wall interacts with disturbances in the shear layer, modifying their frequency modes. According to our previous findings \cite{bhaduri2024effects}, the impinging shock amplifies the KH instability within the shear layer, enhancing mixing and reducing the dominant frequency of the cavity configurations compared to cases without the shock under the same flow conditions and cavity geometry. This section investigates how the impinging shock wave influences the flow field in cavities with varying aspect ratios ($L/D$) subjected to flows of different M$_\infty$.

 As explained above, the periodic feedback loop within the cavity is a key flow characteristic common to all these setups. To study this, we first identify a complete cycle using the time history of the normalized pressure (p/p$_\infty$) recorded by probes placed at the front (P1) and aft walls (P2) of the cavities. Subsequently, we extract normalized density gradient contours($\nabla \rho \cdot (D/\rho_{\infty})$) for the corresponding time instances to highlight the associated flow. Time (t) is normalized by T, defined as the ratio of the cavity length (L) to the freestream velocity (U$_\infty$).
\subsubsection{$L/D$=3}
Figure \ref{fig:4} depicts the temporal variation of the pressure at the front and aft walls of the cavity of $L/D=3$ and M$_\infty$ of 1.71. The time interval t/T = 69.02-70.77 captures a complete oscillation cycle inside the cavity. At the beginning of the cycle, Figures \ref{fig:4}a and \ref{fig:4}b indicate a high pressure at the aft wall, while at the front wall, it is seen to decrease. Around t/T = 70.2–70.3, the pressure at the front wall peaks, while the aft wall pressure dips. Beyond this point, the pressure at the aft wall rises again as it decreases at the front wall.

Figure \ref{fig:5} presents the synthetic schlieren, correlating the pressure variations at the cavity walls with key flow events inside the cavity. At t/T = 69.02, high pressure recorded by the aft wall probe results from the pressure wave (1) generated inside the cavity as mass enters the cavity through the trailing edge (Figure \ref{fig:5}a). The shock (3) impinges on the shear layer (2) at the impinging point (IP), initially predicted at 1.061D from the leading edge (Table \ref{tab:table2}) but observed at 0.406D at the beginning of the cycle due to its interaction with the leading-edge separation shock (4) at point (6). This streamwise distance of 0.406D remains nearly constant throughout the cycle until the last observed instant at t/T = 70.77, which also marks the beginning of the next cycle.

The pressure wave (1), tracked within the blue square in subsequent figures, propagates upstream and impinges on the front wall at t/T = 70.12 (Figure \ref{fig:5}f), causing a pressure peak at the front wall while the aft wall experiences minimal pressure. This low aft-wall pressure induces mass entrainment through the trailing edge, initiating the next oscillation cycle. During this phase, a reattachment shock (7) forms as the shear layer impinges on the aft wall, preceding mass entrainment. The shear layer reflects the shock wave (3) as an expansion wave (5) to maintain pressure balance. Readers can refer to the appendix (Sec \ref{appendix}) for a detailed explanation of this reflection using acoustic impedance analysis.

The alternating compression and expansion near the IP generate strong velocity gradients, amplifying shear and promoting energy transfer from the mean flow to fluctuating fields. This process strengthens the formation of Kelvin-Helmholtz (KH) rolls downstream of the IP. Additionally, the pressure wave inside the cavity enhances the growth of these vortical structures as it travels downstream after reflecting from the front wall, as depicted in Figures \ref{fig:5}(g)–\ref{fig:5}i. The complete cycle, illustrated in Figures \ref{fig:5}(a)–\ref{fig:5}(i), spans an interval of t/T = 1.75.
\begin{figure*}

	\includegraphics[scale=1.4]{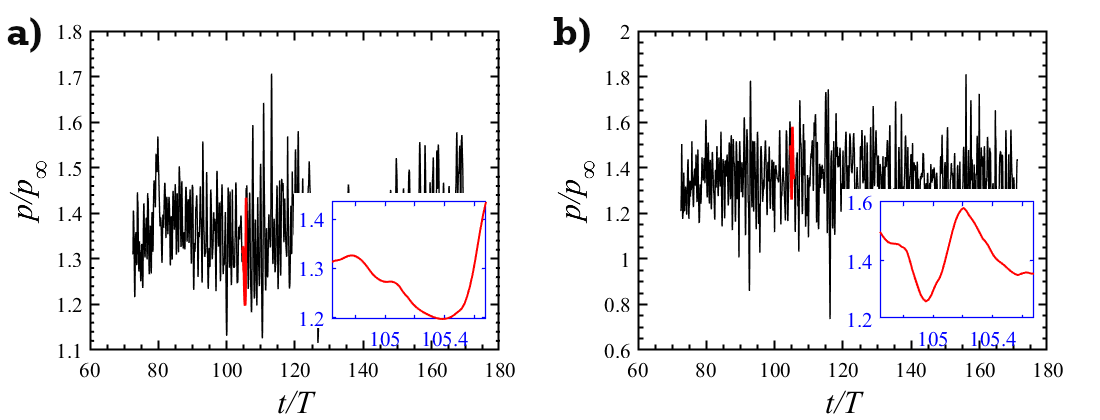}% Here is how to import EPS art
    \centering

    \caption{\label{fig:12} Temporal variation of pressure normalized with the freestream pressure (p/p$_\infty$) at the a) front and b) aft walls of the cavity of L/D = 2 at M$_\infty$ = 1.71. The time is normalized with T (L/U$_\infty$ = 5.513e-5 s).}
   \end{figure*}
    \begin{figure*}

	\includegraphics[scale=0.32]{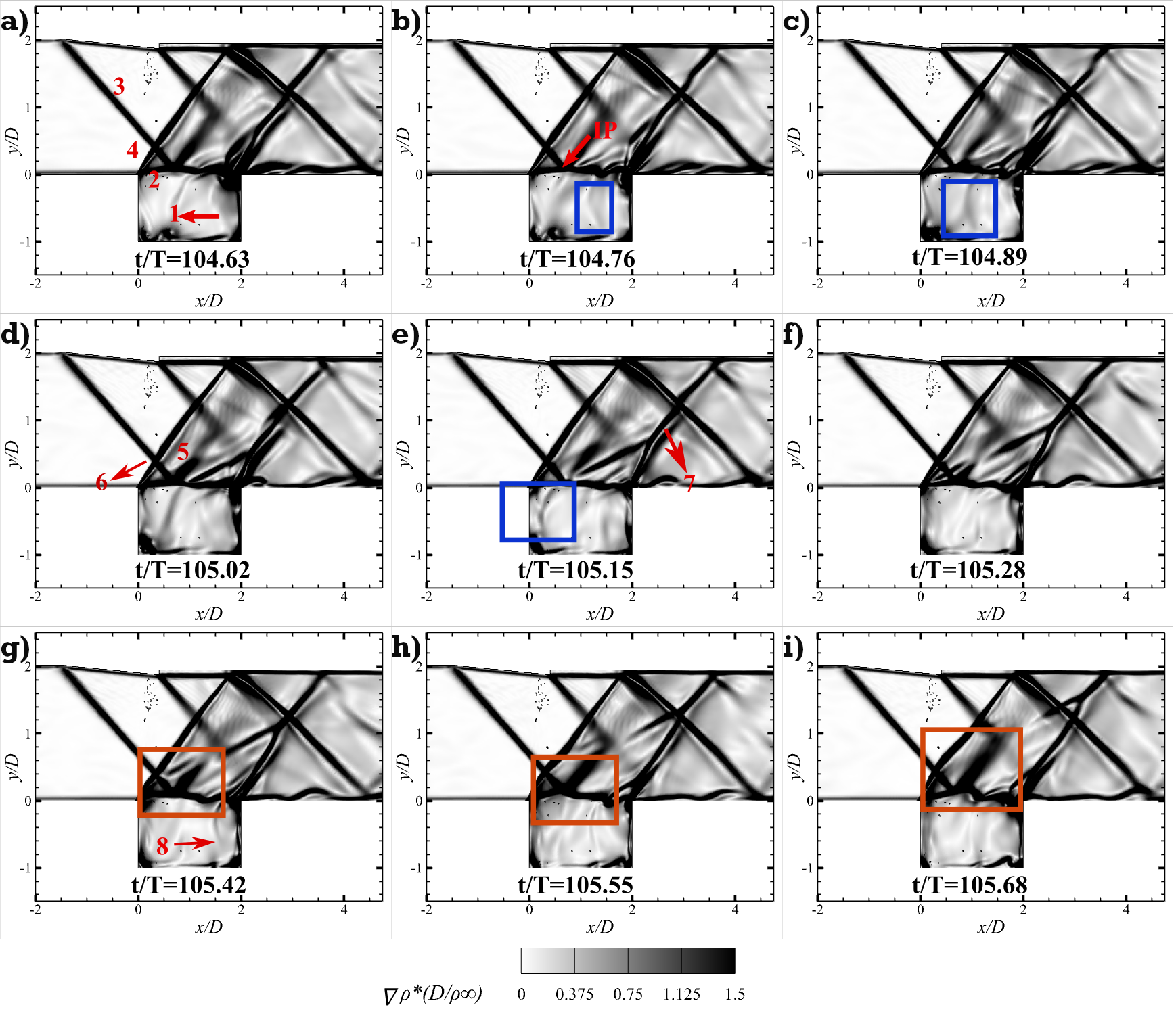}% Here is how to import EPS art
    \centering

    \caption{\label{fig:13} Normalized Density Gradient contour ($\nabla \rho \cdot (D/\rho_{\infty})$) from the time step (t/T) of 104.63 (a) to 105.63 (i) at an interval of 0.131 for the cavity of L/D=2 at M$_\infty$ = 1.71.Key flow features: (1) the upstream traveling wave, (2) the perturbed separating shear layer, (3) the impinging shock, (4) the separating shock wave, (5) the expansion wave,  (6) the interaction points of (3) and (4), (7) reattachment shock, and (8) upstream traveling pressure wave of the next cycle. IP is the impinging point. The blue square marks the upstream traveling pressure wave. The red square demonstrates the shear layer perturbations.}
   \end{figure*}
   
   \begin{figure*}

	\includegraphics[scale=1.4]{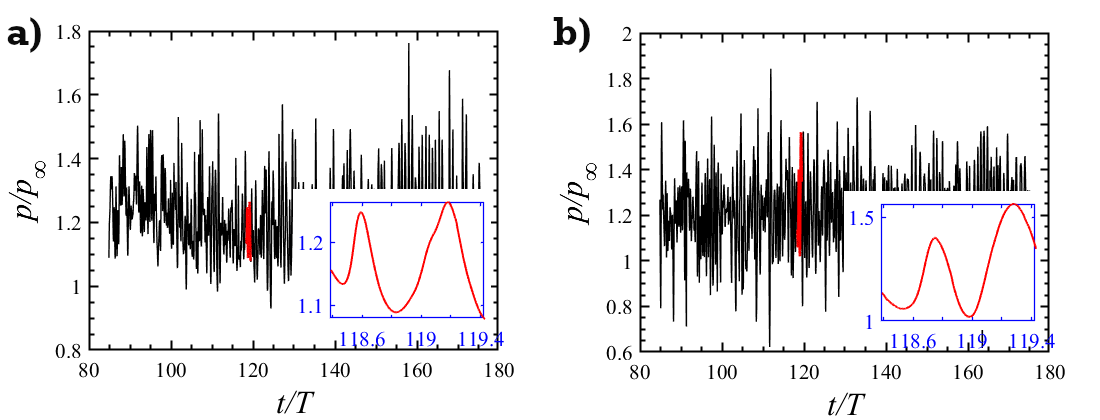}% Here is how to import EPS art
    \centering

    \caption{\label{fig:14} Temporal variation of pressure normalized with the freestream pressure (p/p$_\infty$) at the a) front and b) aft walls of the cavity of L/D=2 at M$_\infty$ = 2. The time is normalized with T (L/U$_\infty$ = 4.714e-5 s).}
   \end{figure*}
    \begin{figure*}

	\includegraphics[scale=0.32]{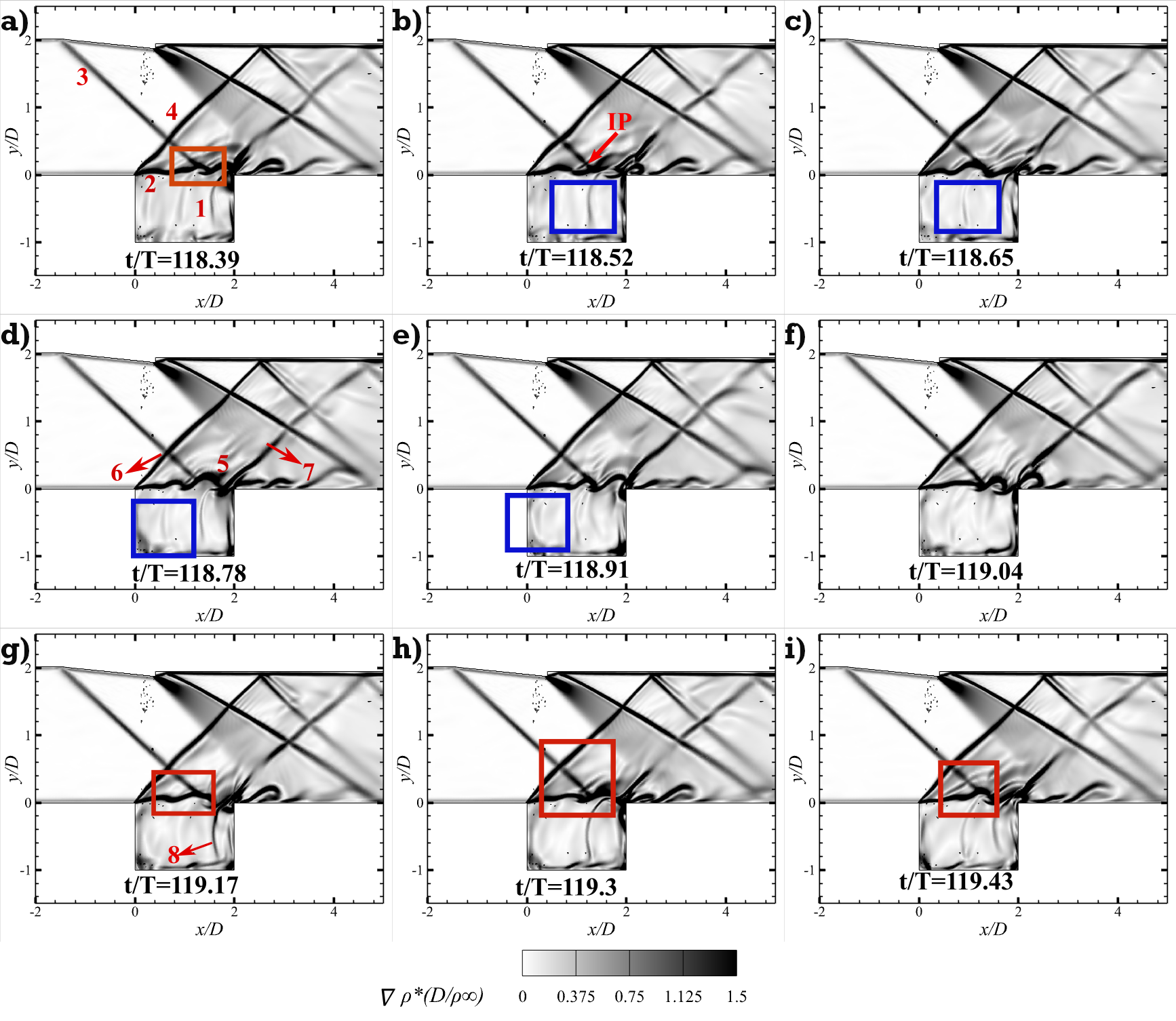}% Here is how to import EPS art
    \centering

    \caption{\label{fig:15} Normalized Density Gradient contour ($\nabla \rho \cdot (D/\rho_{\infty})$) from the time step (t/T) of 118.39 (a) to 119.43 (i) at an interval of 0.13 for the cavity of L/D = 2 at M$_\infty$ = 2. Key flow features: (1) the upstream traveling wave, (2) the perturbed separating shear layer, (3) the impinging shock, (4) the separating shock wave, (5) the expansion wave,  (6) the interaction points of (3) and (4), (7) reattachment shock, and (8) the upstream travelling pressure wave of the next cycle. IP is the impinging point. The blue square marks the upstream traveling pressure wave. The red square demonstrates the KH rolls.}
   \end{figure*}
   
   \begin{figure*}

	\includegraphics[scale=1.4]{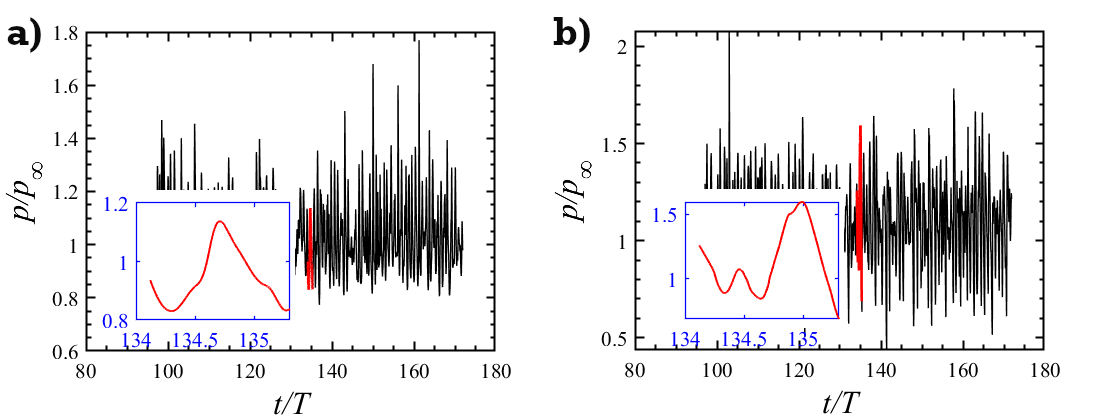}% Here is how to import EPS art
    \centering

    \caption{\label{fig:16} Temporal variation of pressure normalized with the freestream pressure (p/p$_\infty$) at the a) front and b) aft walls of the cavity of L/D=2 at M$_\infty$ = 2.29. The time is normalized with T (L/U$_\infty$ = 4.12e-5 s).}
   \end{figure*}
   \begin{figure*}

	\includegraphics[scale=0.35]{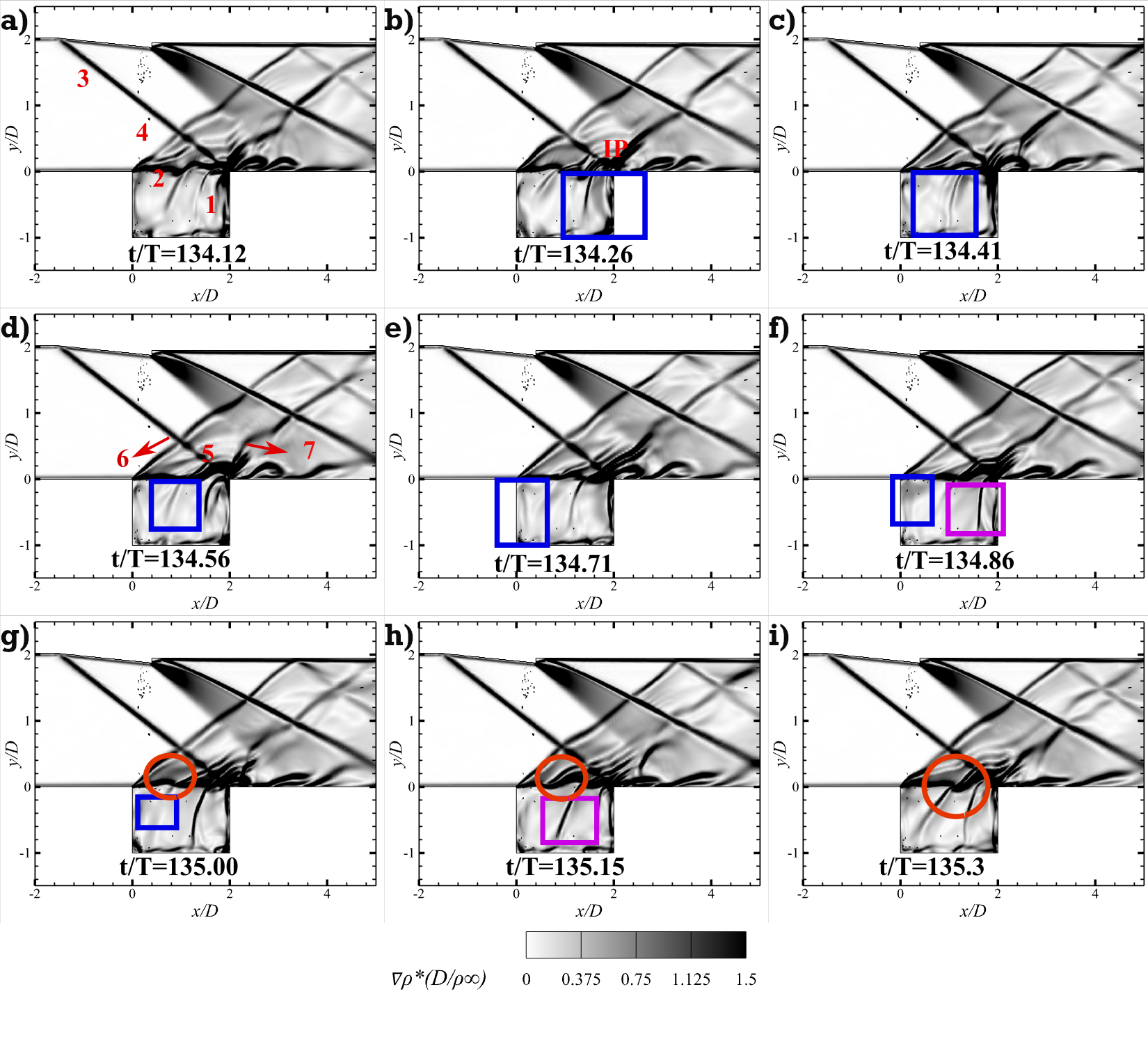}% Here is how to import EPS art
    \centering

    \caption{\label{fig:17} Normalized Density Gradient contour ($\nabla \rho \cdot (D/\rho_{\infty})$)  from the time step (t/T) of 134.12 (a) to 135.3(i) at an interval of 0.131 for the cavity of L/D=2 at M$_\infty$ = 2.29. Key flow features: (1) the upstream traveling wave, (2) the perturbed separating shear layer, (3) the separation shock, (4) the separating shock wave, (5) the expansion wave,  (6) the interaction points of (3) and (4), and (7) reattachment shock. IP is the impinging point. The blue square marks the upstream traveling pressure wave. The red circle demonstrates the disturbances in the shear layer, and the purple square represents the upstream traveling pressure wave of the next cycle.}
   \end{figure*}
     \begin{figure*}

	\includegraphics[scale=1.4]{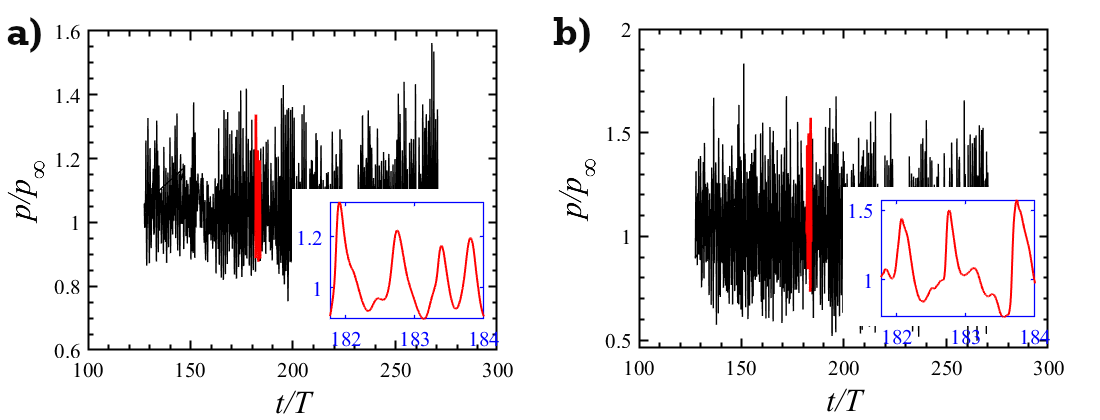}% Here is how to import EPS art
    \centering

    \caption{\label{fig:18} Temporal variation of pressure normalized with the freestream pressure (p/p$_\infty$) at the a) front and b) aft walls of the cavity of L/D=2 at M$_\infty$ = 3. The time is normalized with T (L/U$_\infty$ = 3.138e-5 s).}.
   \end{figure*}
 \begin{figure*}

	\includegraphics[scale=0.32]{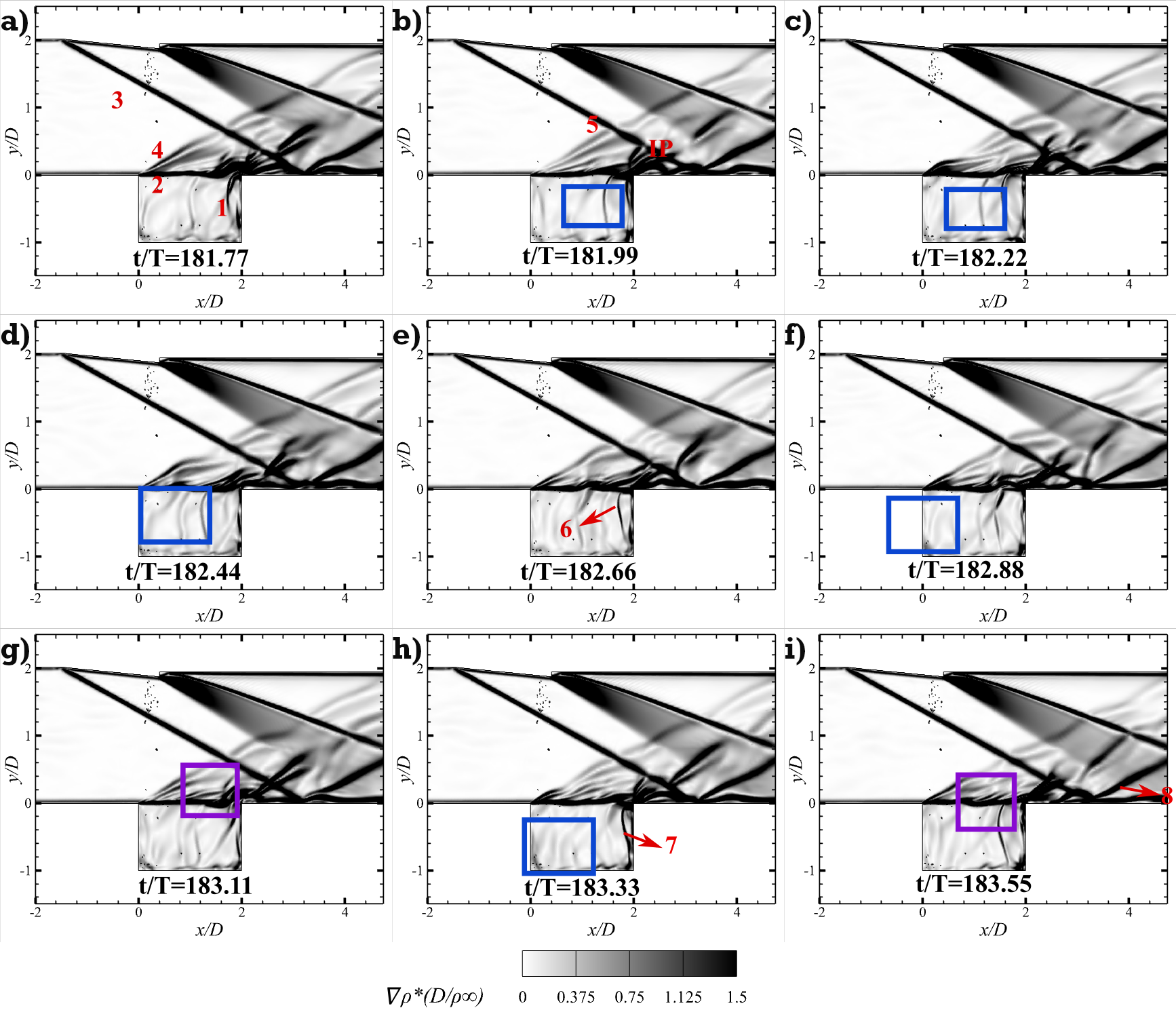}% Here is how to import EPS art
    \centering

    \caption{\label{fig:19} Normalized Density Gradient contour ($\nabla \rho \cdot (D/\rho_{\infty})$) of the cavity without confinement for one complete cycle from the time step (t/T) of 181.77 (a) to 183.55(i) at an interval of 0.222 for the cavity of L/D=2 at M$_\infty$ = 3. Key flow features: (1) the upstream traveling wave, (2) the perturbed separating shear layer, (3) the separation shock, (4) the separating shock wave, (5) the interaction points of (3) and (4) , (6) and (7) the upstream traveling pressure wave of the next cycles, and (8) reflected shock from the surface. IP is the impinging point. The blue square marks the upstream traveling pressure wave. The purple square demonstrates the perturbations in the shear layer.}
   \end{figure*}

Figure \ref{fig:6} depicts the temporal pressure variation for the cavity with $L/D=3$ at M$_\infty$ = 2, showing a similar trend to M$_\infty = 1.71$. At t/T = 87.68, the pressure recorded by the probe on the aft wall is high due to the generation of the pressure wave (1) inside the cavity, caused by mass entrainment through the trailing edge (Fig. \ref{fig:7}a). As the pressure wave moves upstream, aft wall pressure decreases, reaching a minimum beyond t/T = 89, while front wall pressure peaks. Figures \ref{fig:7}(f)–\ref{fig:7}(g) show the pressure wave striking and reflecting from the front wall during this high-pressure phase. Figure \ref{fig:7} g also shows the reattachment shock forming at the trailing edge as the distances impinge on the aft wall. Due to the interactions with the separation shock (4) at point (6), the shock (3) impinges on the shear layer (2) at 0.84D at the beginning of the cycle at t/T=87.68 (0.88D upstream of the predicted IP as seen in Table \ref{tab:table2}). The streamwise distance of 0.84D of the impinging point from the leading edge remains almost similar for the entire cycle, similar to M$_\infty$=1.71. However, due to the high intensity of the oscillations of the shear layer, the distance varies in the longitudinal direction. The stronger shock impinging the shear layer at higher M$_\infty$ leads to greater flow property variations. Steep velocity gradients enhance shear, facilitating energy transfer from the mean flow to fluctuations. This increased energy amplifies disturbances in the shear layer, enhancing fluid entrainment, as seen in the Schlieren images in Figure \ref{fig:7}.  A reattachment shock (7) also forms as the disturbances impinge on the trailing edge. The entire feedback cycle in Figures \ref{fig:7}(a)- \ref{fig:7}(i) lasts for 1.72, slightly faster than in the case of M$_\infty$ = 1.71 due to the increase in speed of the flow.

The impinging shock, in addition to acoustic excitation of the shear layer vortices by the back-and-forth travelling pressure wave inside the cavity, plays a crucial role in sustaining KH instability growth in confined cavities. The shock and expansion waves reflected from the shear layer intensify velocity gradients near the impinging point (IP), facilitating energy transfer to fluctuating fields. However, as the M$_\infty$ increases, the compressibility effects disrupt the nonlinear growth phase of KH instability due to the oscillatory pressure and velocity fields induced by the vortex reversal. This hindrance in the growth of the KH instability also attenuates the shear layer growth for high freestream M$_\infty$ \cite{ragab1989linear,papamoschou1988compressible,sandham1991three}.

At M$_\infty$ = 2.29 (Fig. \ref{fig:9}b), the impinging point (IP) shifts to 1.14D as observed at the beginning of the cycle at t/T = 94.53 (1.188D upstream of the predicted value (Table \ref{tab:table2})). Under these conditions,  amplification of the perturbations in the shear layer is weaker than in lower  M$_\infty$ cases. Energy transfer remains sufficient to induce secondary instabilities, manifesting as vortex rolls and their interactions (Fig. \ref{fig:9}). At approximately t/T = 95, the pressure wave impinges near the front wall, reaching the maximum front wall pressure (Fig. \ref{fig:8}a). Simultaneously, the aft wall pressure also peaks, signaling mass entrainment through the trailing edge and cycle initiation, as highlighted in the black box in Figure \ref{fig:9}c. A reattachment shock (5) is seen as the shear layer impinges the aft wall. The upstream-traveling pressure wave reflected from the front wall further amplifies perturbations in the shear layer via resonance, as observed in the shear layer beyond t/T = 95.03 (Figs. \ref{fig:9}(d)-\ref{fig:9}(h)). The complete feedback loop spans a time interval of t/T = 1.97, making it slower than that observed at lower M$_\infty$.

The upstream shift of the IP from the calculated values as observed at the beginning of the presented feedback cycle,  increases with M$_\infty$ till 2.29 due to the strong shock-shock interactions, which compress the flow more effectively and shift the IP upstream. However, at M$_\infty$=3, the IP shift is reduced to 0.914D, placing it at 2.78D, which is nearer to the trailing edge at t/T=118.81 (Fig. \ref{fig:11}). The impinging shock (3) interacts with the separation shock (4) and an additional wave (5), likely generated by acoustic wave reflection from the front wall (Fig. \ref{fig:11}b), affecting the upstream shift of the IP.

The compressibility effects at a higher M$_\infty$ of 3 suppress the growth of the KH instability due to vortex reversal (Fig. \ref{fig:11}). The shock impinges on the shear layer ahead of the trailing edge, which also disrupts the spatial growth of the vortices. The pressure profiles at the front and aft walls of the cavity reveal intriguing patterns in this case. Beyond t/T = 119.5, the pressure at both walls rises simultaneously (Fig. \ref{fig:10}). The pressure increase at the front wall indicates the impingement of upstream-traveling pressure waves inside the cavity, while mass entrainment through the trailing edge and the initiation of a new cycle elevate the pressure at the aft wall. 
Disturbances in the shear layer are slightly amplified by the oscillatory motion of the pressure wave within the cavity, which appears as minor curvatures in the shear layer from t/T = 119.11 onward (Figs. \ref{fig:11}(c)-\ref{fig:11}(i)). At t/T = 119.11, a downstream-traveling pressure wave perturbs the shear layer, coinciding with the lowest front-wall pressure (Fig. \ref{fig:10}a). This suggests that the pressure wave from the previous cycle has propagated downstream from the leading edge. By 
t/T = 119.5, the front-wall pressure reaches its peak (Fig. \ref{fig:10}a), marking the impingement of the current cycle’s pressure wave on the front wall.
The perturbation in the shear layer intensifies after t/T = 119.57 as it interacts with the reflected pressure wave (Figs. \ref{fig:11}(f)-\ref{fig:11}(i)). These disturbances grow spatially as they convect downstream toward the trailing edge. Around t/T=120, the aft wall pressure increases (Fig. \ref{fig:10}b), signaling the onset of the next cycle, as visualized in the Schlieren images (Fig. \ref{fig:11}i). Some disturbances at the cavity base amplify as the pressure wave interacts with them while traveling downstream (Fig. \ref{fig:11}e). These disturbances, highlighted by a green square, propagate upstream toward the front wall, while the reflected pressure wave moves downstream, perturbing the shear layer between t/T = 119.42 and t/T = 119.87 (Figs. \ref{fig:11}(e)-\ref{fig:11}(h)). Consequently, the front-wall pressure rises again, peaks, and subsequently declines (Fig. \ref{fig:10}a). The feedback cycle is completed in t/T = 1.21, the shortest duration among all M$_\infty$ cases for the cavity of L/D = 3.

\subsubsection{$L/D$=2}
Figure \ref{fig:13} presents synthetic Schlieren images for a cavity with L/D = 2 at 
M$_\infty$ =1.71. The temporal pressure evolution at the front and aft walls (Fig. \ref{fig:12}) corroborates these observations. As the feedback cycle initiates, a pressure wave (1) forms near the aft wall inside the cavity and propagates upstream, reducing aft-wall pressure while increasing front-wall pressure (Figs. \ref{fig:12}(a)-\ref{fig:12}(b)). This wave, marked by a blue square, interacts with the shear layer (2), amplifying disturbances before reflecting from the front wall (Figs. \ref{fig:13}(g)-\ref{fig:13}(i)). At t/T = 105.4, the next feedback cycle begins, marked by high aft-wall pressure, low front-wall pressure, and the emergence of a new pressure wave (8). The feedback cycle completes in t/T = 1.05, which is 0.7 times faster than for L/D = 3 due to the shorter travel distance at the same speed.

 \begin{figure*}

	\includegraphics[scale=1.1]{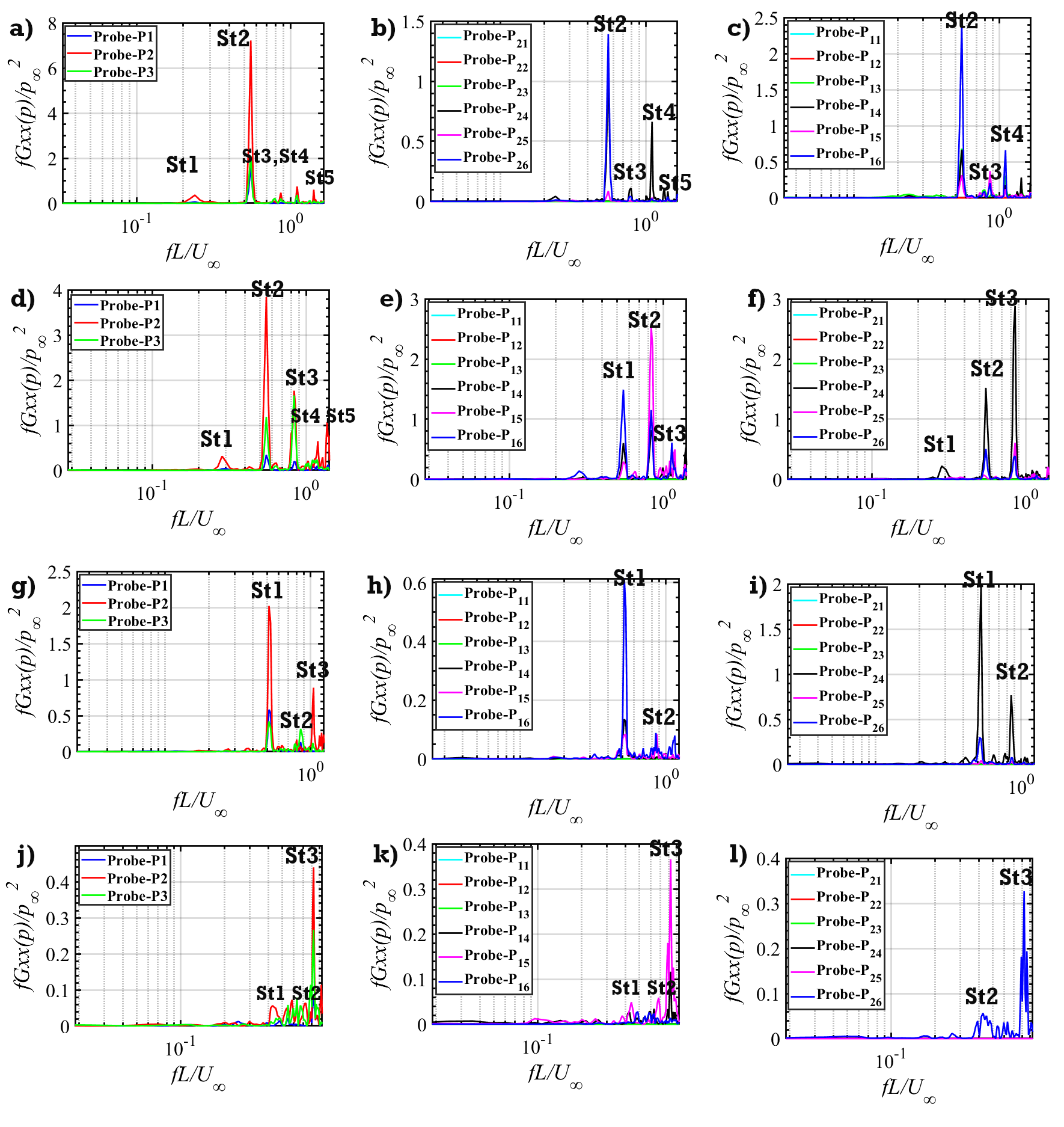}% Here is how to import EPS art
    \centering

    \caption{\label{fig:20} Normalized Power Spectral Density (PSD) (fGxx(p)/$(p_\infty)^2$) vs the Strouhal number (St=fL/$U_\infty$) for cavity with L/D = 3, at: (a)-(c) M$_\infty$ = 1.71,(d)-(f) M$_\infty$ = 2,  (g)-(i) M$_\infty$ = 2.29, (j)-(l) M$_\infty$ = 3.  The first column represents internal probes, while the second and third columns correspond to the probes at probe racks PR1 and PR2.}
   \end{figure*}
\begin{figure*}

	\includegraphics[scale=1.1]{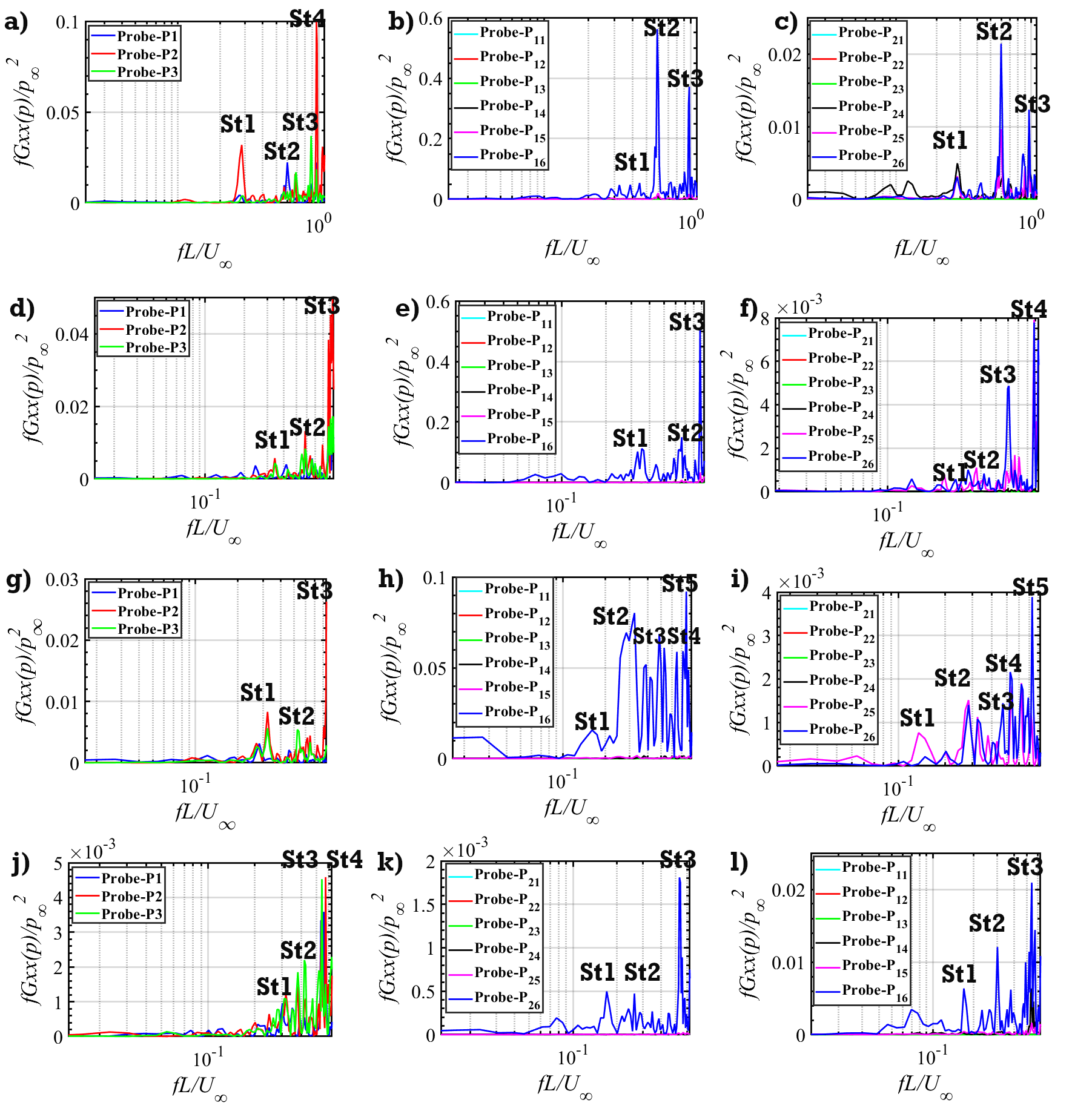}% Here is how to import EPS art
    \centering

    \caption{\label{fig:21} Normalized Power Spectral Density (PSD) (fGxx(p)/$(p_\infty)^2$) vs Strouhal number (St=fL/$U_\infty$) for the cavity with L/D = 2 at : (a)-(c) M$_\infty$ = 1.71, (d)-(f) M$_\infty$ = 2,  (g)-(i) M$_\infty$ = 2.29, (j)-(l) M$_\infty$ = 3.  The first column represents internal probes, while the second and third columns correspond to the probes at probe racks PR1 and PR2.}
   \end{figure*}
   \begin{table*}[ht]
    \centering
    \caption{\label{tab:table3} Comparison of modes obtained from analytical solution and present simulations for $L/D=3$.The dominant Strouhal number and its corresponding frequency are highlighted in bold.}
    \begin{ruledtabular}
    \begin{tabular}{c c c c}
        \textbf{Mach Number} & \textbf{Modes from Analytical Equation} & \textbf{Modes from Simulations} & \textbf{Dominant Frequency} \\
        \hline
        1.71 & 0.24, 0.56, 0.883, 1.204, 1.52 & 0.24, \textbf{0.554}, 0.86, 1.14, 1.4 & \textbf{6.798kHz}\\
        2    & 0.231, 0.539, 0.847, 1.155, 1.46 & 0.283, \textbf{0.55}, 0.834, 1.189, 1.36 & \textbf{7.67kHz}\\
        2.29 & 0.22, 0.521, 0.82, 1.12         & \textbf{0.521}, 0.852, 1.16 & \textbf{8.44kHz} \\
        3    & 0.21, 0.49, 0.78         & 0.44, 0.67, \textbf{0.82} & \textbf{17.424kHz} \\
        %4    & 0.202, 0.472                    & 0.207, 0.44, 0.547, \textbf{0.65}& \textbf{18.413kHz} \\
    \end{tabular}
    \end{ruledtabular}
\end{table*}
% Table 1
\begin{table*}[ht]
    \centering
    \caption{\label{tab:table4} Comparison of modes obtained from analytical solution and present simulations for $L/D=2$. The dominant Strouhal number and its corresponding frequency are highlighted in bold.}
    \begin{ruledtabular}
    \begin{tabular}{c c c c}
        \textbf{Mach Number} & \textbf{Modes from Analytical Equation} & \textbf{Modes from Simulations} & \textbf{Dominant Frequency}\\
        \hline
       1.71 & 0.24, 0.56, 0.88 & 0.284, 0.59, 0.88,\textbf{0.95} & \textbf{17.24kHz} \\
        2    & 0.23, 0.54, 0.85 &  0.24, 0.58,\textbf{0.94}& \textbf{19.94kHz} \\
        2.29 & 0.224, 0.523, 0.823        & 0.323,0.523,\textbf{0.823} & \textbf{19.97kHz} \\
        3    & 0.2114, 0.5         & 0.24, 0.53, \textbf{0.57} &\textbf{18.164kHz} \\
       % 4    & 0.202, 0.472                    & 0.207, 0.44, 0.547,\textbf{0.65} \\
    \end{tabular}
    \end{ruledtabular}
\end{table*}

\begin{figure*}

	\includegraphics[scale=0.45]{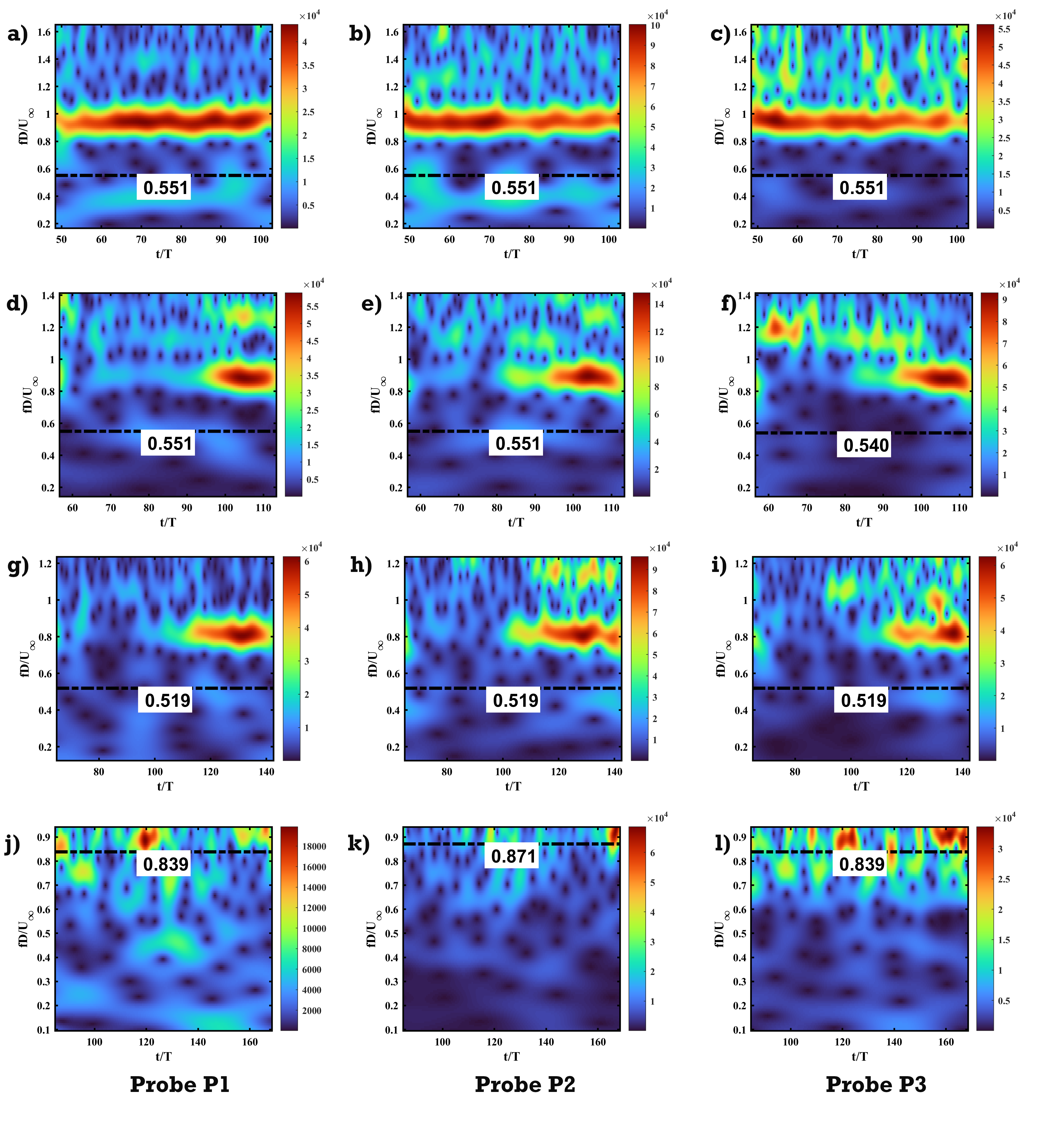}% Here is how to import EPS art
    \centering

    \caption{\label{fig:22} Continuous Wavelet Transform (CWT) of probes for a cavity with L/D = 3, showing the Strouhal number (St=fL/$U_\infty$) over normalized time (t/T) for :(a)-(c) M$_\infty$ = 1.71,(d)-(f) M$_\infty$ = 2,  (g)-(i) M$_\infty$ = 2.29, (j)-(l) M$_\infty$ = 3.}
   \end{figure*}
  
  \begin{figure*}

	\includegraphics[scale=0.45]{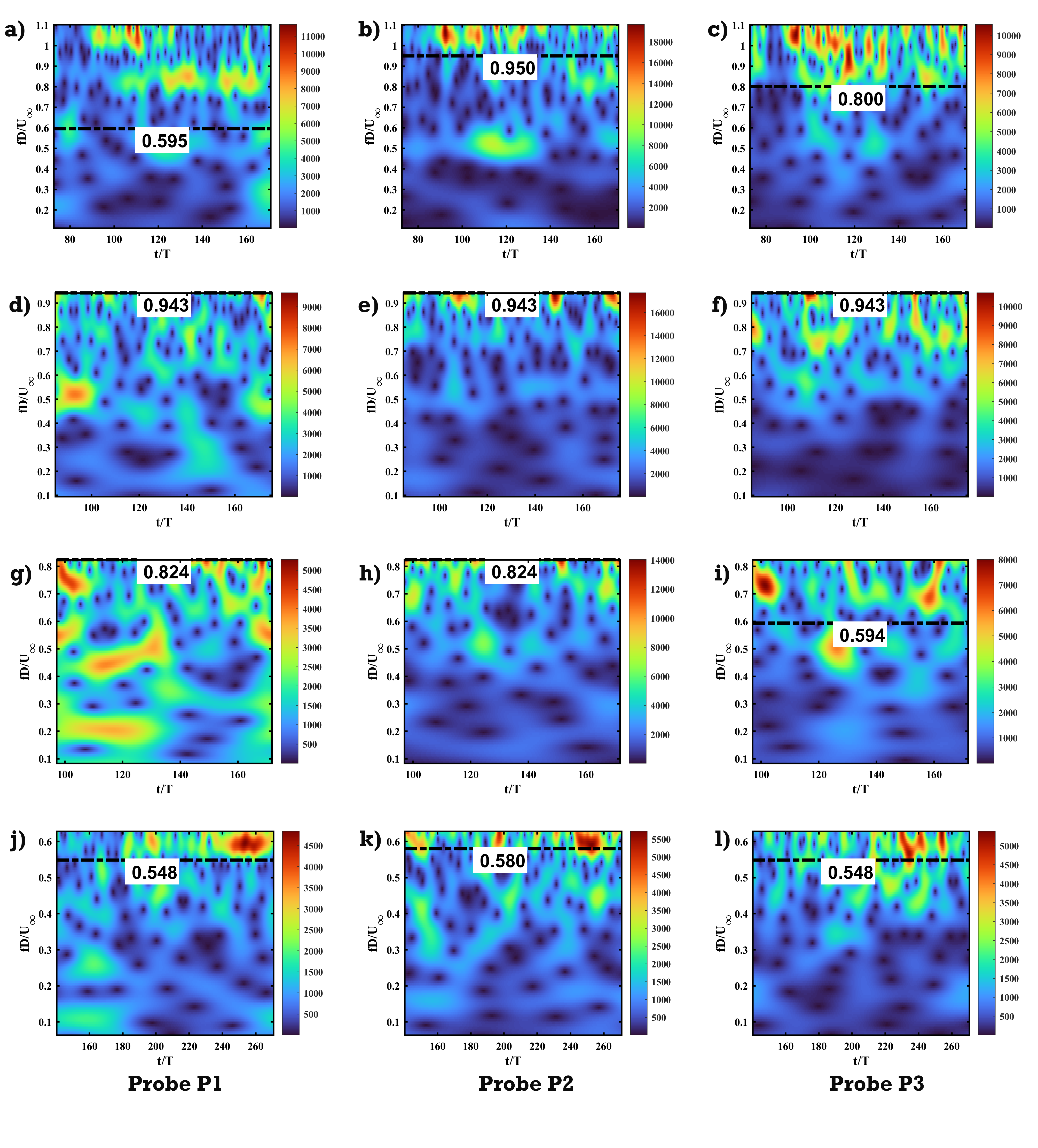}% Here is how to import EPS art
    \centering

    \caption{\label{fig:23} Continuous Wavelet Transform (CWT) of probes for a cavity with L/D = 2, showing the Strouhal number (St=fL/$U_\infty$) over normalized time (t/T) for :(a)-(c) M$_\infty$ = 1.71,(d)-(f)M$_\infty$ = 2,  (g)-(i) M$_\infty$ = 2.29, (j)-(l) M$_\infty$ = 3.}
   \end{figure*}
The impinging shock (3) impinges on the shear layer (2) after its interaction with the leading edge separation shock (4), generating an expansion fan (5) downstream of the impingement point (IP). However, unlike the L/D=3 case, the development of the KH rolls is hindered. The IP appears at 0.5964D from the leading edge at t/T 104.63, which is 0.19D downstream than in the L/D = 3 case, suggesting that the reflected shock, formed by the interaction of the impinging shock (3) and the separation shock (4), impinges on a more developed shear layer. Consequently, the expansion fan is weaker, and the velocity gradient does not reach the threshold required for sustaining the KH roll formation. Additionally, the spatial growth of KH instability is affected, as disturbances travel a shorter distance before reaching the trailing edge than in the L/D = 3 cases, further inhibiting their development.

At M$_\infty$ = 2, the shock (3) impinges the shear layer (2) 0.33D downstream compared to the L/D = 3 case, as captured in the 
 beginning of the feedback cycle at t/T = 118.39. (Fig. \ref{fig:15}). Likewise, in the case of the cavity configuration of L/D=3, the streamwise distance of the IP from the leading edge remains almost unchanged throughout this cycle till t/T = 119.43, which is also a part of the next cycle. The velocity gradient near the IP reaches the threshold needed for KH roll sustainance in this case, in spite of the impingement of the shock in a more developed shear layer.  However, the growth of the perturbation remains lower than in the L/D = 3 case, indicating reduced energy transfer from the mean flow and insufficient distance for spatial growth. Numerical Schlieren images (Fig. \ref{fig:15}) confirm that the feedback cycle completes in t/T = 1.04, which is faster than its L/D = 3 counterpart. The pressure evolution at the cavity walls (Figure 14) exhibits a reduced time lag between cycles, as evidenced by two peaks per cycle. During 
$t/T=119-119.4$, a new pressure wave (8) forms at the trailing edge (Fig. \ref{fig:15}g) and propagates upstream as the reflected wave from the front wall moves downstream.

At M$_\infty$ = 2.29, the impinging shock (3) impinges on the shear layer (2) at a streamwise distance of 
1.742D, which is 0.6D  downstream than in the L/D = 3 case (Fig.\ref{fig:17}b) at the beginning of the described feedback loop at t/T = 134.12. The increased velocity gradient across the IP amplifies the perturbations, manifesting itself as a higher shear layer perturbation ahead of the aft wall (Fig. \ref{fig:17}). The evolution of the pressure on the front wall (Fig. \ref{fig:16} a) shows that the upstream traveling pressure wave (1) reaches the front wall between
$t/T=134-135$, where it reflects and further amplifies shear layer disturbances (Fig. \ref{fig:17}e). Simultaneously, another pressure wave (purple square) forms at the aft wall (Fig.\ref{fig:17}f), propagating upstream while the preceding wave (blue square) moves downstream. Numerical Schlieren images (Fig. \ref{fig:17}) confirm a feedback cycle duration of t/T = 1.176, which is shorter than for L/D = 3.

As M$_\infty$ increases to 3, the IP shifts to 2.88D, moving 
0.1D further downstream than in the L/D = 3 case, as we have seen at the beginning of the feedback cycle at t/T = 181.77. This location lies beyond the core cavity flow, rendering the impinging shock (3) ineffective in altering cavity dynamics. The duration of the feedback loop is t/T = 1.78 as reflected from the pressure peaks (Fig. \ref{fig:16}) and Schlieren images (Fig. \ref{fig:17}). Unlike in the case of L/D=3, the feedback loop takes a longer time to complete in this configuration.

Flow visualizations across different L/D ratios indicate that, for a given M$_\infty$, feedback cycle time is shorter for L/D=2 due to the reduced travel distance at the same speed. For both the configurations, the feedback cycle duration decreases with increasing M$_\infty$ up to 2 but then increases at M$_\infty$ =2.29. At M$_\infty$ $\ge$2.29, the cycle time decreases again for L/D = 3 but continues increasing for L/D = 2. 

Apart from the cavity aspect ratio, shock impingement location and Kelvin-Helmholtz (KH) instability are the two dominant flow features that influence feedback dynamics. For the cavity of L/D = 3, at M$_\infty$ = 3, the shock impinges on the shear layer just ahead of the cavity aft wall, increasing the energy of the perturbations in the shear layer before they impinge on the aft wall. The increased compressibility effect at M$_\infty$, the location of the impingement point, which affects the spatial growth of the KH instability, and also the increased speed with M$_\infty$, have a cumulative effect on the faster duration of the feedback loop. Similarly, in the case of the cavity with the L/D = 2 case at the M$_\infty$ = 2.29, the IP is ahead of the aft wall, enhancing the energy transfer to the perturbations in the shear layer. However, at M$_\infty$ = 3, the shock impinges on the flow downstream of the cavity's trailing edge, having little influence on the flow inside the cavity. Therefore, for the cavity of L/D=2, the increase in time to complete the feedback loop is influenced more by the amplification of the perturbations in the shear layer upstream of the aft wall than the speed of the flow.  

The increase in time to complete the feedback loop, despite the increasing velocity at M$_\infty$ = 2.29, for both the configurations, presents an intriguing phenomenon. The increase in duration between M$_\infty$ = 2 and 2.29 is smaller for the L/D = 2 case than for L/D = 3. This difference is attributed to the shock impingement location and the spatial distance over which perturbations develop before reaching the aft wall.

For all M$_\infty$ values, the IP shifts upstream due to the interaction between the impinging shock and the leading-edge separation shock, generating a reflected shock that impinges on the shear layer. Despite identical flow conditions, this upstream shift is smaller for L/D=2. We also observe that the KH instability is more prominent in L/D = 3 cases, with compressibility effects reducing the intensity of the KH rolls with the increase in M$_\infty$. The presence of KH rolls indicates a higher energy transfer from the mean flow to the fluctuating field, implying an intense mixing of flow properties. In L/D = 2 cases, the shear layer is more perturbed for M$_\infty$ = 2 and 2.29, while for L/D = 3, it is more perturbed for M$_\infty$ = 1.71 and 2. Therefore, the L/D ratio of the cavity, flow conditions, position of shock wave impingement on the shear layer, and the intensity of the shear layer disturbance all have cumulative effects on the flow features of the cavity. In the subsequent sections, we analyze quantitatively the effects of these factors on cavity oscillations.

 \begin{figure*}

	\includegraphics[scale=0.38]{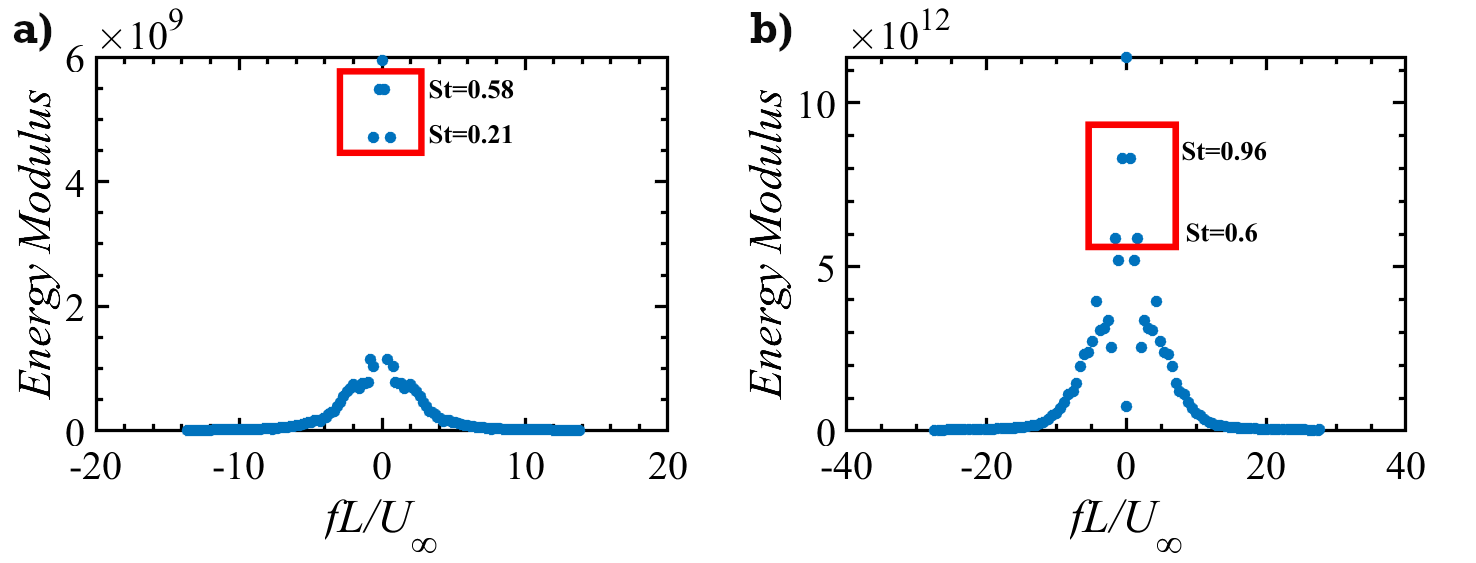}% Here is how to import EPS art
    \centering

    \caption{\label{fig:24} Energy associated with the Strouhal numbers for cavity configuration of (a) L/D=3 and (b) L/D=2 at M$_\infty$=1.71.}
   \end{figure*}
   \begin{figure*}

	\includegraphics[scale=0.3]{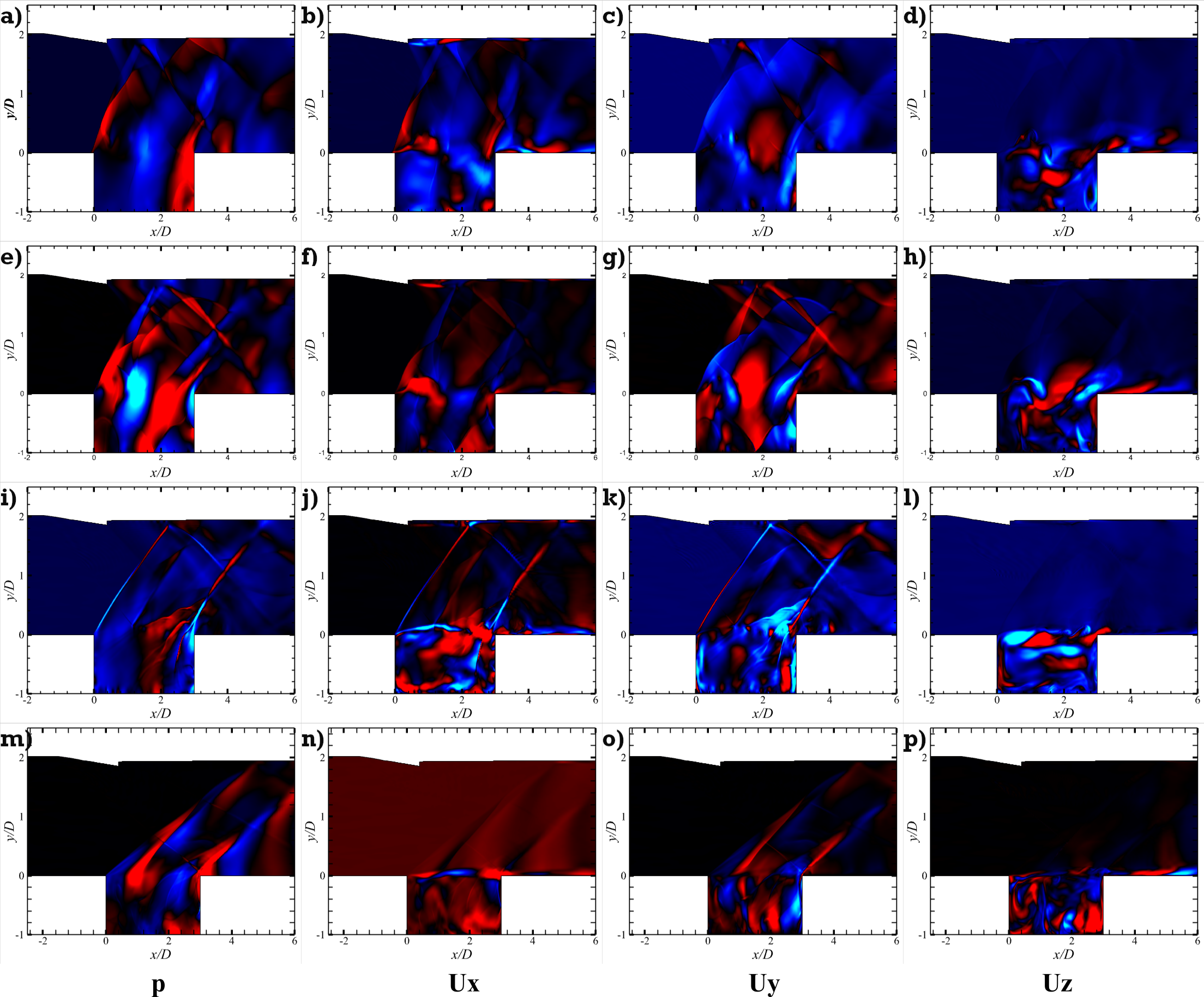}% Here is how to import EPS art
    \centering

    \caption{\label{fig:25} Pressure and velocity fields corresponding to the dynamic mode related to the dominant frequency for the cavity with L/D = 3 at (a)-(d) M$_\infty$ = 1.71, (e)-(h) M$_\infty$ = 2 (i)-(l) M$_\infty$ = 2.29 and (m)-(p) M$_\infty$ = 3.}
   \end{figure*}

   \begin{figure*}

	\includegraphics[scale=0.32]{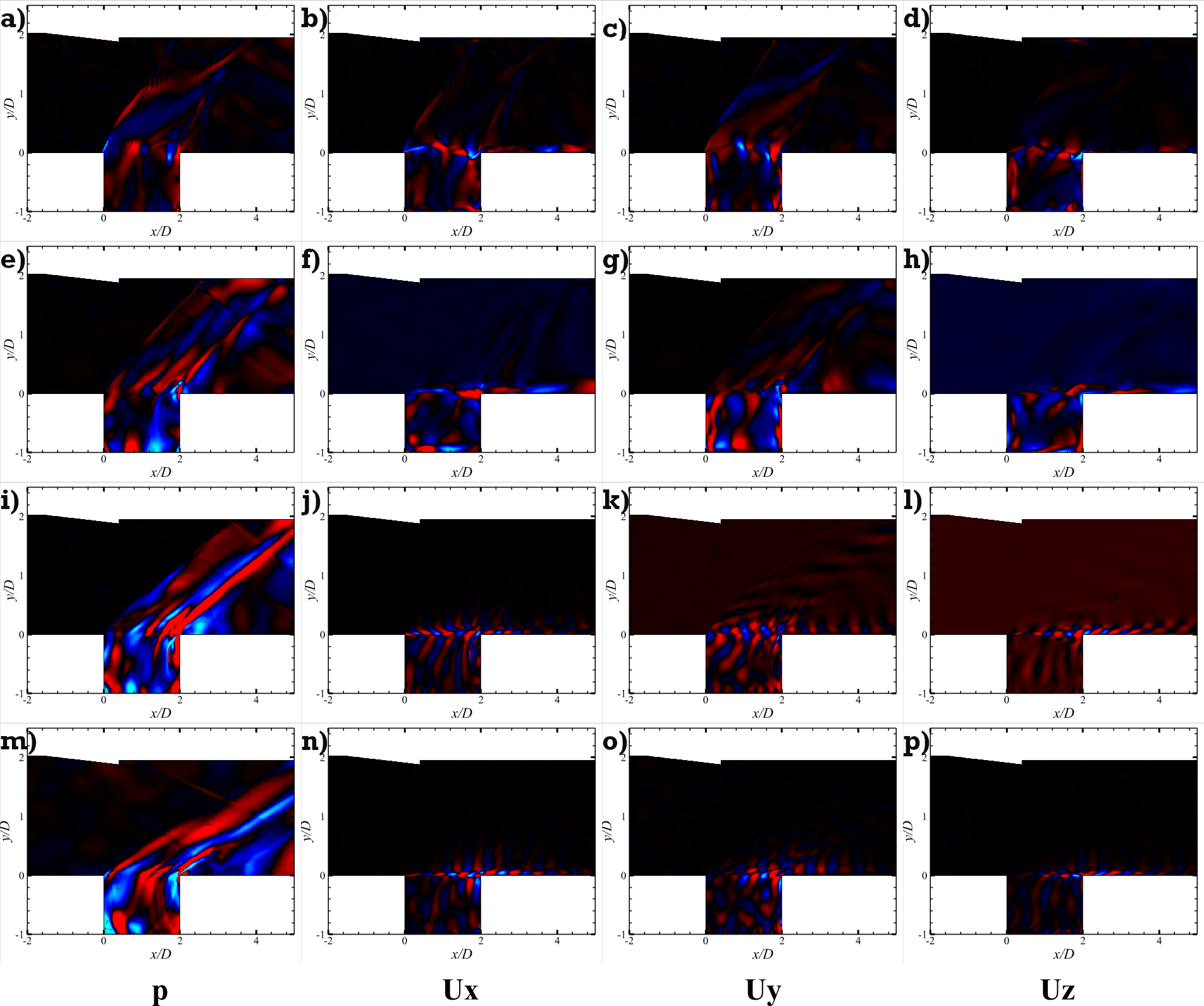}% Here is how to import EPS art
    \centering

    \caption{\label{fig:26} Pressure and velocity fields corresponding to the dynamic mode related to the dominant frequency for the cavity with L/D = 2 at (a)-(d) M$_\infty$ = 1.71, (e)-(h) M$_\infty$ = 2 (i)-(l) M$_\infty$ = 2.29 and (m)-(p) M$_\infty$ = 3.}
   \end{figure*}
    
\subsection{Spectral Analysis} \label{section:SA}
  
In this section, we determine the frequency content of the system, the associated energy of these modes, and the time evolution of these frequencies. Probes placed at the cavity walls and near the shear layer, record data at an interval of 2e-08 s corresponding to a 50 MHz sampling rate. This ensures compliance with the Nyquist criterion and enables a broad frequency resolution. Each probe collects sufficient samples for all the configurations, yielding a 200 Hz data resolution for an accurate spectral analysis. Power Spectral Density (PSD) is computed using the 'pwelch' algorithm with a Hanning window and 50\% overlap to analyze energy distribution in the unsteady pressure signal.
 Continuous Wavelet Transformation (CWT) analysis determines the temporal evolution of frequencies present in the system. The wavelet transformation resolves frequency and time with greater precision than the Fourier transformation, offering deeper insights into the frequencies present in the system.
Readers can refer to Chui \cite{Chui1992}, Torrence \cite{Torrence1998}, and Debnath \cite{Debnath2014} for further details on wavelet transformation.
\subsubsection{Power Spectral Density Analysis}
Figures \ref{fig:20} and \ref{fig:21} present the PSD for cavity configurations L/D=3 and L/D=2 across all M$_\infty$.The y-axis shows the power spectral density, normalized with the ratio of f/p$_\infty$$^2$, while the x-axis represents the Strouhal number (St= fL/U$_\infty$). The first column displays spectral data from probes placed along the internal cavity walls (as mentioned in section \ref{sec:level2}), while the subsequent columns show data from six probes at probe racks PR1 and PR2. The analysis reveals the following key observations:
\begin{itemize}
\item{Figures \ref{fig:20}(a)–\ref{fig:20}(c) show that for M$_\infty$ = 1.71 for cavtiy of L/D = 3, the second mode (St2= 0.554) dominates across all probes. At M$_\infty$ = 2, the second mode (St2 = 0.55) remains dominant at the internal probes and PR1 near the shear layer, while the third mode (St3 = 0.85) dominates the PR2 (Figs. \ref{fig:20}(d)–\ref{fig:20}(f)). For M$_\infty$ = 2.29, the first mode (St1 = 0.521) (Figs. \ref{fig:20}(g)–\ref{fig:20}(i)) and for M$_\infty$ =3, the third mode (St3 = 0.82) emerge as dominant (Figs. \ref{fig:20}(j)–\ref{fig:20}(l)). }

\item{Figures \ref{fig:21}(a)–\ref{fig:21}(c) show that for L/D = 2 at M$_\infty$ = 1.71, the aft wall probe (P2) captures the dominant mode at St4 = 0.96, while the front wall probe (P1) detects St2 = 0.59, and the base probe (P3) captures St3 = 0.88. The probe racks register St2 = 0.59 as dominant. As M$_\infty$ increases to 2 and 2.29, the dominant Strouhal number decreases to 0.94 (Figs.\ref{fig:21}(d)–\ref{fig:21}(f)) and 0.823 (Figs. \ref{fig:21}(g)–\ref{fig:21}(i)), respectively. At M$_\infty$ = 3, the dominant St further drops to 0.56 (Figs. \ref{fig:21}(j)–\ref{fig:21}(l)). }

\item{For both the configurations, internal probes consistently capture the dominant Strouhal number from probe P2 at the aft wall. This location corresponds to the formation of the pressure wave, which is an integral part of the feedback loop inside the cavity. Additionally, the power spectral density in the internal probes decreases with M$_\infty$ for both cases.}

\item{In PR1 for L/D=3, probe $P_{16}$ (located at 2D from the leading edge) captures the dominant frequency for M$_\infty$ =1.71 and M$_\infty$=2.29, while $P_{15}$ (at 1D from the leading edge) does so for other cases. In PR2, probe $P_{24}$ (at 0.5D from the leading edge) registers the dominant frequency for 
M$_\infty$ =2 and 2.29, whereas $P_{26}$ captures it for 
M$_\infty$ =1.71 and M$_\infty$ =3. Additionally, for M$_\infty$ =1.71, energy is higher in PR1 at y=0.1D; for M$_\infty$ = 2 and 3, probes in PR1 and PR2 exhibit similar energy content, while for M$_\infty$ = 2.29, PR2 shows higher energy.}

\item{For L/D = 2, probe $P_{n6}$ in both probe racks consistently captures the dominant frequency, with significantly higher energy in PR1 than PR2.At M$_\infty$ = 2.29, $P_{25}$ and $P_{26}$ in PR2 register multiple lower frequencies (St = 0.162–0.52), indicating higher unsteadiness near the aft edge.}

\item{The dominant Strouhal numbers from the spectral analysis closely match with the time taken for the feedback cycle in the flow visualization(Section \ref{section:flow}) across all M$_\infty$ (Figs. \ref{fig:4}–\ref{fig:19}).}
\item{Probes placed along the inner walls of the cavity and at the probes racks near the shear layer capture higher energy for L/D = 3 than for L/D = 2. This indicates the higher energy flow to the fluctuations from the mean field in cavities with L/D = 3 and justifies the intense perturbations in the respective shear layer as seen in the numerical schlieren in Sec \ref{section:flow}.}
\end{itemize}
\subsubsection{Comparison with the analytical solution}

Tables \ref{tab:table3} and \ref{tab:table4} compare the Strouhal number of the modes obtained from the modified Rossiter's formula (equation \ref{eq:rossiter})and the PSD analysis for $L/D=3$ and $L/D=2$, respectively. The dominant Strouhal number for each configuration is highlighted in bold and corresponds to the probes placed at the aft wall (P2). The frequency is normalized using $L/U_\infty$ to determine the Strouhal number (St).

\begin{equation}
\frac{f L}{U_{\infty}} = \frac{n - \alpha}{\left( M_{\infty} \left( 1 + \frac{\gamma - 1}{2} M_{\infty}^2 \right)^{-0.5} + \frac{1}{\kappa} \right)}
\label{eq:rossiter}
\end{equation}

where, L is the characteristic length, $\kappa$=0.47 and $\alpha$=0.25 as suggested in the literature by Heller \cite{heller1971flow}.
 \begin{table*}
    \caption{\label{tab:table5}Modes obtained from the Dynamic Mode Decomposition for all cavity configurations}
    \begin{ruledtabular}
    \begin{tabular}{c|c|c}  
        \textbf{M$_\infty$} & \textbf{High Energy Mode} & \textbf{Low Energy Mode} \\
        \hline
        \multicolumn{3}{c}{\textbf{L/D = 3}} \\
        \hline
        1.71  & 0.58  & 0.21  \\
        2.00  & 0.55  & -  \\
        2.29  & 0.50  & 0.24  \\
        3.00  & 0.85  & -  \\
        \hline
        \multicolumn{3}{c}{\textbf{L/D = 2}} \\
        \hline
        1.71  & 0.96  & 0.60  \\
        2.00  & 0.92  & 0.54  \\
        2.29  & 0.84  & 0.54  \\
        3.00  & 0.59  & -  \\
    \end{tabular}
    \end{ruledtabular}
\end{table*}

For $L/D=3$, as M$_\infty$ increases, the dominant Strouhal number decreases when the shock impinges on the shear layer upstream of the trailing edge. However, when the impingement point nears the trailing edge (M$_\infty$ =3), the dominant Strouhal number rises. The dominant frequency increases with M$_\infty$, but since St is inversely proportional to $U_\infty$, their combined effect results in a decreasing trend in St for M$_\infty$ between 1.71 and 2.29. However, between M$_\infty$ =2.29 and 3, the dominant frequency increases by approximately 106\%, while $U_\infty$ increases by 31.2\%, leading to an increase in the dominant St for M$_\infty$ =3. Additionally, the dominant St for M$_\infty$ =3 slightly exceeds the predicted Rossiter modes. The shock impinging upstream of the trailing edge reflects as an expansion wave, generating strong property gradients in the flow near the trailing edge, which re-enters the cavity. The speed of the flow is also high in the case of this M$_\infty$. Moreover, increased compressibility effects may influence the upstream speed of sound, a critical factor in Rossiter’s relation, contributing to deviations from analytical predictions.  
%These factors collectively explain the higher frequency observed at M$_\infty$ = 3.

For $L/D=2$, the dominant frequency increases with increasing M$_\infty$. The frequencies for each M$_\infty$ in this case are higher than those for the $L/D=3$ cavity. The shorter feedback loop completion time $L/D=2$ leads to higher oscillation frequencies. Flow visualization further reveals lesser shear layer disturbances in this configuration. A reduced cavity length and increased oscillation frequency weaken the acoustic-vortex resonance due to frequency mismatch, preventing effective disturbance amplification. 

Our previous study \cite{bhaduri2024effects} indicates that impinging shock enhances KH instability, increasing energy transfer from the mean flow to fluctuations. As a result,  a significant part of the total energy is spent in the energy exchanges, and less energy remains available to sustain the feedback loop, leading to lower oscillation frequencies in cases with impinging shocks. The normalized power spectral density values in Figures \ref{fig:20} and \ref{fig:21} confirm that the energy content in the probes at the cavity walls and near the shear layer in L/D = 2 cavities is lower than in L/D = 3. Since KH instability develops spatially, reducing the cavity length restricts the shear layer's evolution, thereby limiting the disturbance growth. This results in reduced energy transfer from the mean flow to fluctuations, forming a weaker shear layer with less prominent KH rolls. As the feedback loop components are interdependent, a weaker shear layer produces a lower-intensity separation shock. This leads to a downstream shift in the impingement point (IP) compared to L/D = 3, reducing the effectiveness of shock-induced KH amplification.

Shock impingement contributes to frequencies higher than analytical predictions at \( M_\infty = 1.71 \) and \( M_\infty = 2 \) for the cavity with \( L/D = 2 \). The dominant frequency at \( M_\infty = 3 \) in this configuration is lower than that at \( M_\infty = 2.29 \). At \( M_\infty = 3 \), the impingement point moves beyond the trailing edge, ceasing to influence the core flow. This suggests that the impinging shock increases the frequency at lower \( M_\infty \), where it interacts with the shear layer. However, further investigation is required to confirm this effect, making it a subject for future research.
\subsubsection{Wavelet Analysis}
Figures \ref{fig:22} and \ref{fig:23} show the energy distribution of frequencies in terms of the Strouhal number for flow structures captured by the probes on the front wall (Probe P1), aft wall (Probe P2), and cavity floor (Probe P3) for cavities with $L/D=3$ and $L/D=2$, respectively. The figures highlight the St, which has the highest energy throughout the entire time duration.  

The value of the dominant St remains the same for all the probes for M$_\infty$ = 1.71-2.29 for the $L/D=3$ cavity throughout the entire duration, as seen in the first three rows of Figure \ref{fig:22}. This trend is in agreement with the PSD analysis (Figs. \ref{fig:20}a,\ref{fig:20}d, \ref{fig:20}g) and flow visualization data (Figs. \ref{fig:5},\ref{fig:7},\ref{fig:9}). At M$_\infty$=3, a slight variation in the dominant St value is seen between the probes, however, the values are near those obtained in the PSD (Fig. \ref{fig:20}j)  and from the flow visualisation data (Fig. \ref{fig:11}). Figure \ref{fig:22} also shows high-energy zones, represented as red blotches in the spectra, corresponding to St greater than the dominant ones. These zones persist at all probe locations throughout the duration at M$_\infty=1.71$. As M$_\infty$ increases, their intensity diminishes, and they become more intermittent. These high-energy frequencies likely result from interactions between the oscillating shear layer and secondary flow structures in the turbulent compressible flow, including small-scale turbulence, higher-order harmonics of the dominant instability, and shock-shear layer interactions. However, they do not represent the dominant frequency of the system, as they are intermittent and localized rather than persisting throughout the entire cycle. The dominant frequency remains the one observed in PSD, wavelet analysis, and flow visualization, as it governs the primary instability mechanism and feedback loop driving the cavity oscillations. As M$_\infty$ increases, the shear layer aligns more parallel to the mean flow, reducing its interactions with vortices and other structures. This alignment weakens secondary instabilities and limits the amplification of high-frequency fluctuations, further confirming that these zones are not the dominant oscillatory mode of the cavity flow.

Figure \ref{fig:23} demonstrates that for L/D=2, the dominant Strouhal number at Probe P2, located at the aft wall of the cavity, matches the value obtained from the PSD analysis for all M$_\infty$ (Figs. \ref{fig:21}a,\ref{fig:21}d, \ref{fig:21}g, \ref{fig:21}j).  At M$_\infty=1.71$, figures \ref{fig:23}a and \ref{fig:23}c indicate dominant Strouhal numbers of 0.595 and 0.8 at the front wall (Probe P1), and cavity floor (Probe P3), respectively. These values correspond to St2 and St3 in the PSD analysis (Figure \ref{fig:21}a). Similarly, for probe P3 on the cavity floor (Fig. \ref{fig:23} i) at M$_\infty=2.29$, the dominant St is 0.594, which closely aligns with St2 (Figure \ref{fig:21}g). For M$_\infty=3$, Figures \ref{fig:23}j and \ref{fig:23}l reveal that Probe P1 on the front wall and Probe P3 at the cavity floor exhibit a dominant Strouhal number of 0.548, which is comparable to St3 from the PSD analysis (Figure \ref{fig:21}j). The cavity with L/D=2 exhibits weaker high-energy zones, indicating a reduced interaction between the suppressed oscillating shear layer and the secondary flow structures.

The frequencies and their respective flow characteristics are further examined using reduced-order modelling in the section \ref{section:DMD}.

\subsection{Dynamic Mode Decomposition} \label{section:DMD}
The synthetic schlieren visualizes the entire flow field of the cavities in Section \ref{section:flow}. Spectral analysis then identifies frequencies associated with cavity oscillations and nonlinear interactions. This section presents the Dynamic Mode Decomposition (DMD) analysis applied to the pressure and velocity fields obtained from LES simulations, to correlate the dominant frequency modes with the flow features. The methodology utilizes parallel QR decomposition and Singular Value Decomposition (SVD) to extract the dynamic modes present in the system \cite{sayadi2016parallel,soni2019modal,arya2021effect,bhatia2019numerical,sharma2024investigation,arya2023acoustic,kutz2016,brunton2016}.The pressure and velocity fields from the LES simulations are recorded at intervals of 2e-5s. For the cases under study, a frequency resolution of 200 Hz and a target maximum frequency of 20 kHz are considered, requiring a total of 200 snapshots. These snapshots are arranged in a matrix. The QR decomposition of this matrix generates an orthogonal matrix (O) and an upper triangular matrix (R). SVD is then applied to R to obtain eigenvalues and eigenvectors. The imaginary part of the eigenvectors reveals the frequency content, while the real part indicates whether a mode is growing, decaying, or oscillatory. The eigenvalues quantify the energy content of each frequency mode.

Figure \ref{fig:24} depicts the energy distribution across various frequency modes, with the x-axis representing frequency in terms of the Strouhal number (St) for cavities with $L/D=3$ and $L/D=2$, at M$_\infty$=1.71. Table \ref{tab:table5} provides the details of the dynamic mode obtained in the frequency range of 2kHz-20kHz for all the cavity configurations in this study. 

Consistent with our previous study \cite{bhaduri2024effects}, we extract the pressure and velocity fields associated with the mode with the highest energy content as seen in figure \ref{fig:24}, which is nearly equal to the dominant St obtained from the flow visualization and spectral analysis. DMD analysis further confirms that the higher mode $St=0.58$ for the cavity of L/D = 3 at M$_\infty$ = 1.71 corresponds to the feedback loop, evident from high pressure near the aft wall (Fig. \ref{fig:25}a). The velocity field reveals large-scale structures, with the streamwise velocity peaking downstream of the leading edge (Fig. \ref{fig:25}b), then decreasing, while the lateral component follows an opposite trend (Fig. \ref{fig:25}c), indicating shear layer perturbations. Flow visualization shows vortex roll-up downstream of the impingement point, captured by the velocity field at $St=0.58$. The spanwise velocity contours further illustrate KH rolls (Fig. \ref{fig:25}d). For $M_\infty=2.29$ (Fig. \ref{fig:25}i), the higher mode, $St=0.5$, corresponds to the feedback loop, as confirmed by the elevated pressure near the aft wall. The streamwise and lateral velocity components are out of phase (Figure \ref{fig:25}(j)-\ref{fig:25}(k)), with the streamwise velocity peaking near the impingement point. The spanwise velocity contour reveals an organized shear layer with anti-symmetric vortex structures (Fig. \ref{fig:25}l). For $M_\infty=2$ and $M_\infty=3$, only one mode falls within the considered frequency range. These modes correspond to the dominant St associated with the feedback loop, as identified in previous analyses. The velocity field features large-scale structures, with streamwise and lateral velocity components out of phase and elevated pressure near the aft wall (Figs.\ref{fig:25}(e)-\ref{fig:25}(g) and \ref{fig:25}(m)-\ref{fig:25}(o)), respectively. However, at $M_\infty=3$, KH instability is suppressed, leading to a less chaotic spanwise velocity field (Fig. \ref{fig:25}p). In contrast, for $M_\infty=2$, the spanwise velocity contour (Fig.\ref{fig:25}h) reveals KH rolls near the leading edge.

For cavity with $L/D=2$ at $M_\infty=1.71$–$2.29$, two modes emerge. The mode with the higher energy content corresponds to a Strouhal number close to values obtained from flow visualization and spectral analysis. The associated pressure and velocity fields exhibit large-scale structures, and the feedback loop, resembles that of the $L/D=3$ case (Figs. \ref{fig:26}(a)-\ref{fig:26}(d),\ref{fig:26}(e)-\ref{fig:26}(h), and\ref{fig:26} (i)-\ref{fig:26}(l)). The spanwise velocity contours indicate shear layer perturbations (Figs. \ref{fig:26}(d),(h) and (l)). The KH rolls are less prominent or absent in the $L/D=2$ case compared to $L/D=3$. At $M_\infty=3$, the $L/D=2$ cavity exhibits only one mode at $St=0.59$ (Table \ref{tab:table5}), aligning with the feedback loop, as indicated by high pressure near the aft wall (Fig. \ref{fig:26}m). The streamwise and lateral velocities remain out of phase (Figs. \ref{fig:26}(n)-\ref{fig:26}(o)). Since KH instability is absent, the spanwise velocity field exhibits minimal structures (Fig.\ref{fig:26}p).

%\FloatBarrier
\section{\label{sec:level4}Conclusion}
In this work, we examine how flow characteristics change under different geometrical and flow conditions in a supersonic open cavity confined by a top wall at a constant deflection angle of $2.29^\circ$. The main conclusions are:
In the L/D = 3 cavity, enhanced Kelvin Helmholtz (KH) instability increases energy transfer to shear layer perturbations. The impinging shock reflects as an expansion fan from the shear layer, inducing high-velocity gradients near the impingement point that generate spatially growing KH rolls. Acoustic-vortex resonance further amplifies coherent structures in the shear layer. Spectral analysis reveals that higher M$_\infty$ leads to increased dominant frequencies due to the higher speed of the flow. However, KH instability moderates this rise by dissipating energy to the vortices.  As Mach number (M$_\infty$)increases, compressibility effects intensify, weakening KH instability. This effect is particularly evident between M$_\infty$ = 2.29 and 3, where compressibility suppresses KH instability and the increase in the speed of the flow causes a sharper frequency increase.  Dynamic Mode Decomposition (DMD) further validates this by associating the dominant mode with specific pressure and velocity fields. The cavity’s L/D ratio significantly affects flow dynamics and oscillation frequencies. Spectral analysis shows that for a given M$_\infty$, the oscillation frequency is higher for L/D = 2 than for L/D = 3 due to a shorter feedback loop. Flow visualization reveals that in the L/D = 2 cavity, the reduced distance before impingement of the shear layer on the cavity's aft wall limits KH instability growth compared to L/D = 3. The shock impinges further downstream in the shear layer of the shorter cavity, interacting with a more developed shear layer and producing a weaker expansion fan with lower velocity gradients. This weakens KH instability, reduces energy transfer to coherent structures, and increases the energy available for the feedback loop. The lesser energy content of the probes placed at the walls of the cavity and the shear layer in the cavity of L/D =2  from the PSD analysis indicates the lesser energy exchange between the various components of the cavity flow. A sharp frequency rise occurs between M$_\infty$ = 1.71 and 2, followed by a smaller increment between M$_\infty$ = 2 and 2.29 due to enhanced shear layer perturbations.

However, instead of the expected increase, an unanticipated frequency drop between M$_\infty$ = 2.29 and 3 at L/D = 2 and an increase in the time to complete the feedback loop for M$_\infty$ = 2.29 for both the cavity configurations indicate a complex role of the impinging shock, necessitating additional research.

\begin{acknowledgments}
The authors acknowledge the National Supercomputing Mission (NSM) for computational resources on 'PARAM Sanganak' at IIT Kanpur, supported by MeitY and DST, Government of India. They also thank the IIT Kanpur Computer Center for additional resources.
\end{acknowledgments}

\section*{Conflict of Interest}
The authors have no conflict of interest to disclose.

\section*{Data Availability}
The data that support the findings of this study are available from the corresponding author upon reasonable request.
\section{Appendixes}
\subsection{Nomenclature}
\begin{tabular}{ll}
%\textbf{Symbol} & \textbf{Description} \\
%\hline
D   & Depth of the cavity \\
L   & Length of the cavity \\
$\beta$ & Shock angle \\
$\theta$ & Deflection angle \\
$H_i$ & Height of the inlet \\
$H_o$ & Height of the outlet \\
$p_\infty$ & Freestream pressure \\
$U_\infty$ & Freestream velocity \\
$M_\infty$ & Freestream Mach number \\
$<p>$ & Average pressure \\
s & Radial distance \\
$\rho_\infty$ & Freestream density \\
$\nabla\rho$ & Gradient of density  \\
t & Time \\
T & Reference time \\
f & Frequency \\
$\frac{fL}{U_\infty}$ & Strouhal number (St) \\
Gxx(p) & Power Spectral Density \\
\end{tabular}
\begin{figure*}[t]

	\includegraphics[scale=0.4]{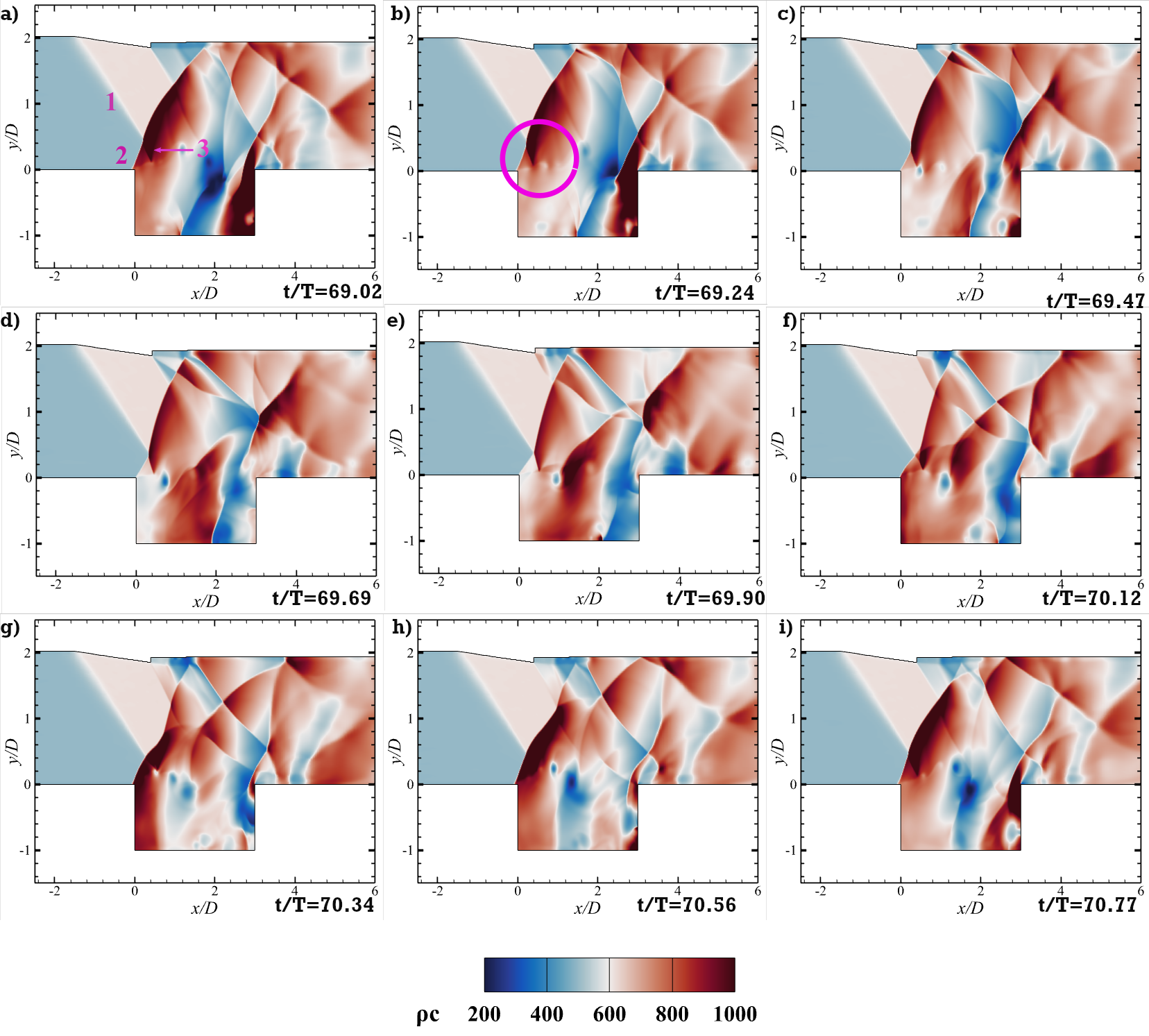}% Here is how to import EPS art
    \centering

    \caption{\label{fig:27}The acoustic impedance contour for the cavity configuration at $L/D=3$ at M$_\infty$=1.71. The feature marked as (1) represents the impinging shock, which first interacts with the leading-edge separation shock (2) before striking the shear layer. Upon interaction, the shear layer reflects the shock as an expansion wave (3). The region of decreasing impedance, highlighted by the purple circle, facilitates the expansion process across (3). }
   \end{figure*}
  
   \begin{figure*}[t]

	\includegraphics[scale=0.4]{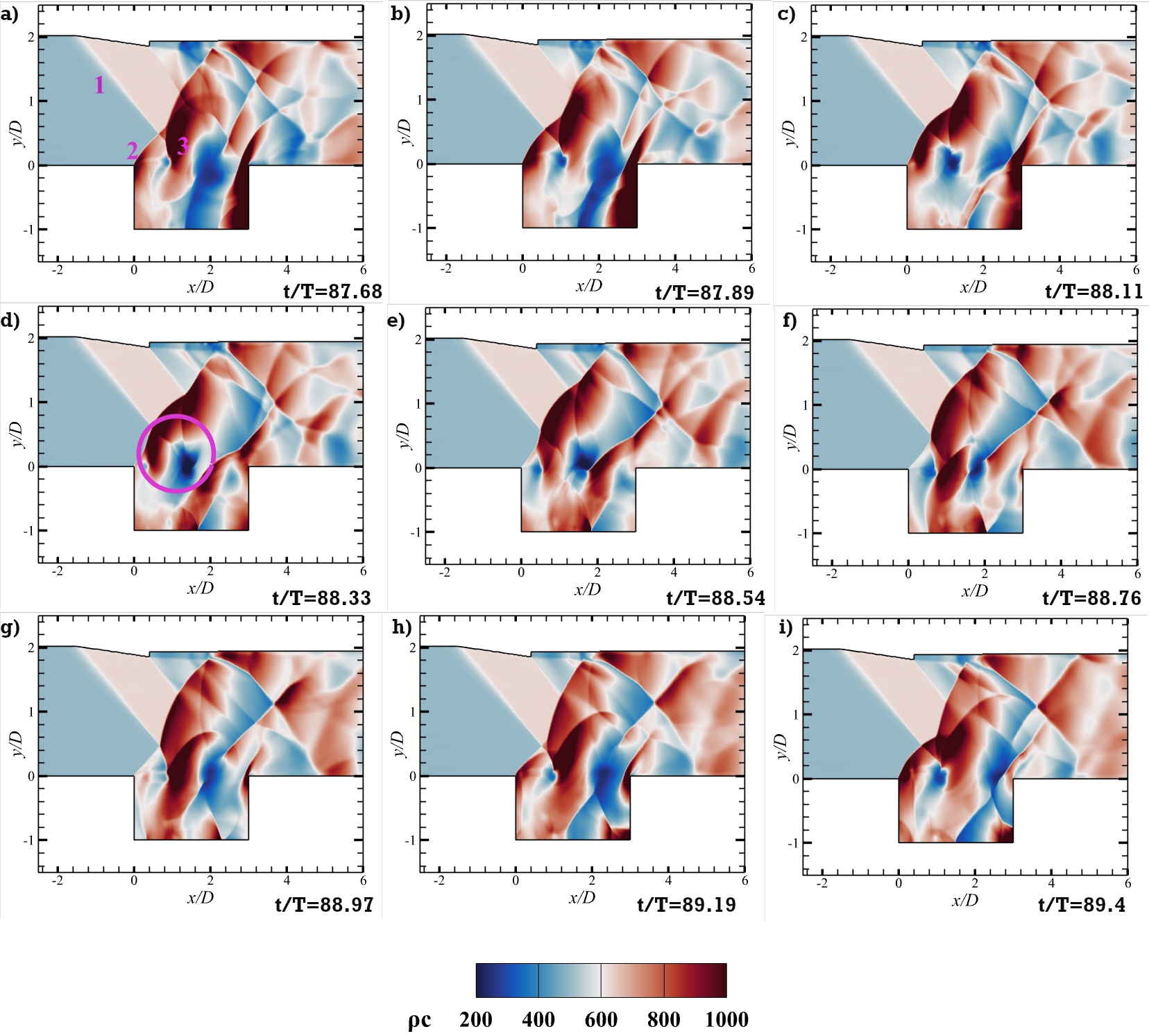}% Here is how to import EPS art
    \centering

    \caption{\label{fig:28}The acoustic impedance contour for the cavity configuration at $L/D=3$ at M$_\infty$=2. The feature marked as (1) represents the impinging shock, which first interacts with the leading-edge separation shock (2) before striking the shear layer.  The shear layer reflects the shock as an expansion wave (3). The region of decreasing impedance, highlighted by the purple circle, facilitates the expansion process across (3).}
   \end{figure*}
   \begin{figure*}

	\includegraphics[scale=0.4]{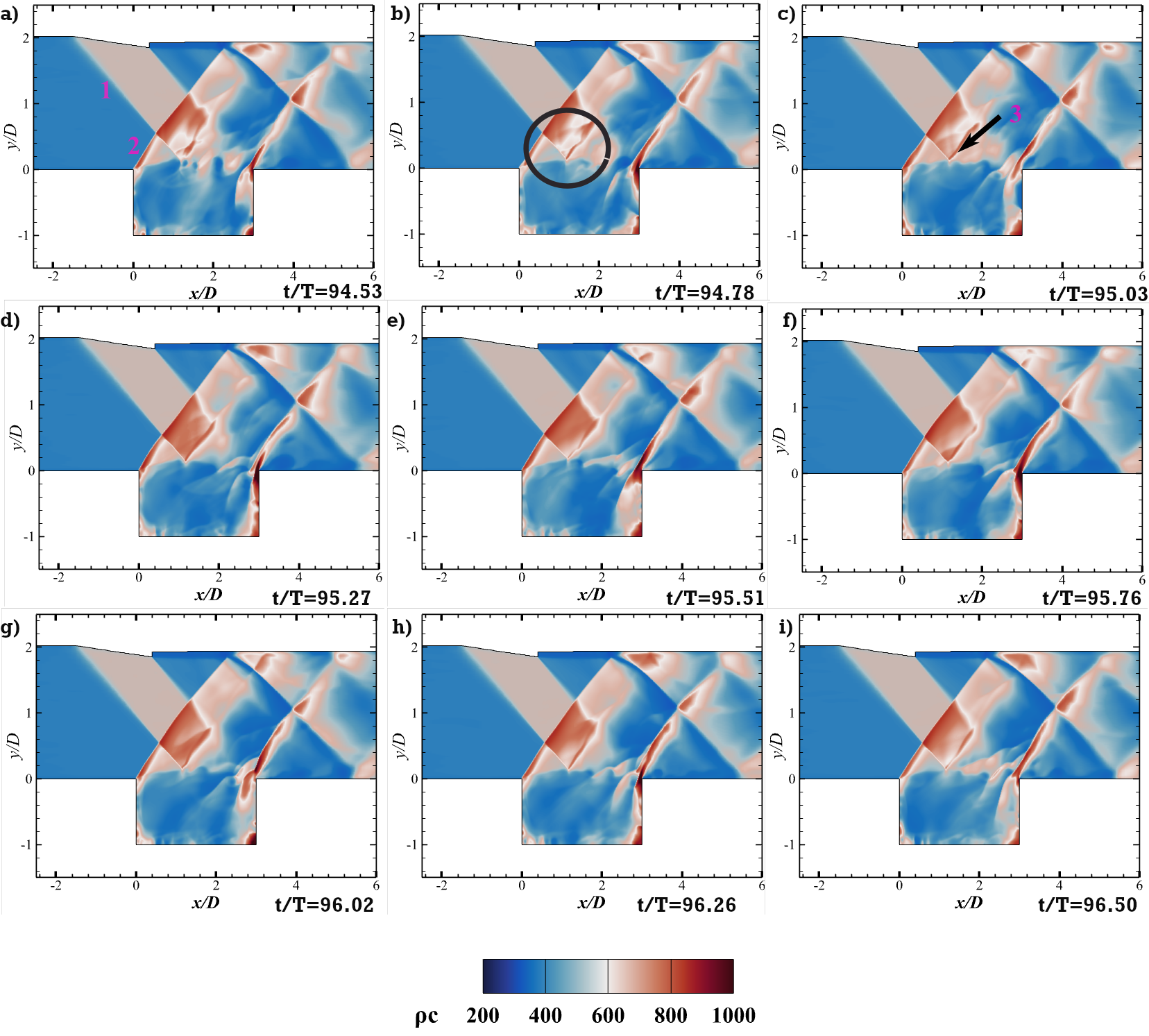}% Here is how to import EPS art
    \centering

    \caption{\label{fig:29} The acoustic impedance contour for the cavity configuration at $L/D=3$ at M$_\infty$=2.29. The feature marked as (1) represents the impinging shock, which interacts with the separation shock (2)  at leading edge before impinging on the shear layer. Upon interaction, the shear layer reflects the shock as an expansion wave (3). The region of decreasing impedance, highlighted by the black circle, facilitates the expansion process, supporting the flow divergence across (3).}
   \end{figure*}
   \begin{figure*}

	\includegraphics[scale=0.4]{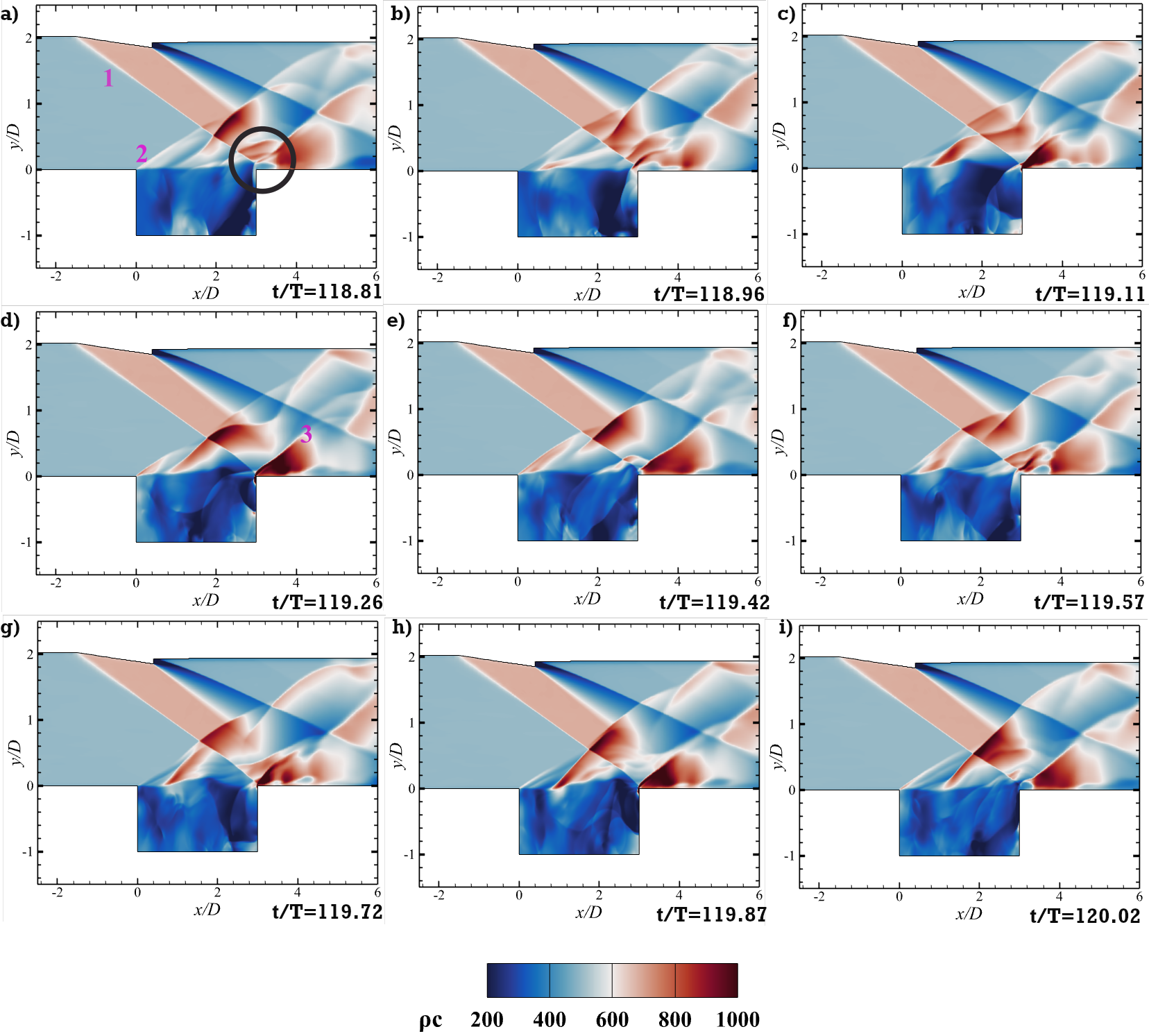}% Here is how to import EPS art
    \centering

    \caption{\label{fig:30}The acoustic impedance contour for the cavity configuration at $L/D=3$ at M$_\infty$=3. The feature marked as (1) represents the impinging shock, which first interacts with the leading-edge shock (2) before striking the shear layer. The flow impinges the aft wall, forming the reattachment shock (3). The region of decreasing impedance is highlighted by the black circle, followed by the increased impedance due to the reattachment shock.  }
   \end{figure*}
   
\subsection{Reflection of Compression Waves as Expansion Waves: An Acoustic Impedance Perspective} \label{appendix}

We analyze the reflection of the compression waves from the shear layer as expansion waves in terms of the acoustic impedance. Acoustic impedance ($\rho$*c) where $\rho$ is density and c is the local speed of sound) characterizes a medium’s resistance to acoustic wave propagation. It plays a crucial role in determining how compression waves interact with flow structures. When a compression wave encounters a region of lower impedance, part of it reflects as an expansion wave to satisfy pressure continuity. Conversely, when it meets a higher impedance region, it strengthens or reflects as another compression wave. 

Figures \ref{fig:27}–\ref{fig:30} present the acoustic impedance contours for the cavity configuration at L/D=3 across all Mach numbers under study. The shock originated at the deflection corner of the top wall, interacts with the leading-edge separation shock, and impinges on the shear layer. In the cavities, the shear layer separates the high-speed freestream from the recirculating cavity flow.
Behind the shock, the flow pressure increases, as indicated by the contours in Figures \ref{fig:27}–\ref{fig:29} for 
M$_\infty$=1.71, 2, and 2.29, respectively. Since the cavity region has a lower overall pressure and density compared to the freestream, the shear layer acts as a lower-impedance interface. When the shock impinges, the pressure difference across the shear layer prevents it from transmitting another compression wave. Instead, to maintain pressure balance, the shear layer reflects the shock as an expansion fan, evident from the sudden decrease in the impedance. For M$_\infty$=3, the shock impinges on the shear layer near the aft wall. As disturbances approach the aft wall, a reattachment shock forms, increasing the local impedance. In this case, the stronger shock system disrupts the usual reflection mechanism, and the expansion fan is less prominently captured.
   \begin{figure*}

	\includegraphics[scale=0.4]{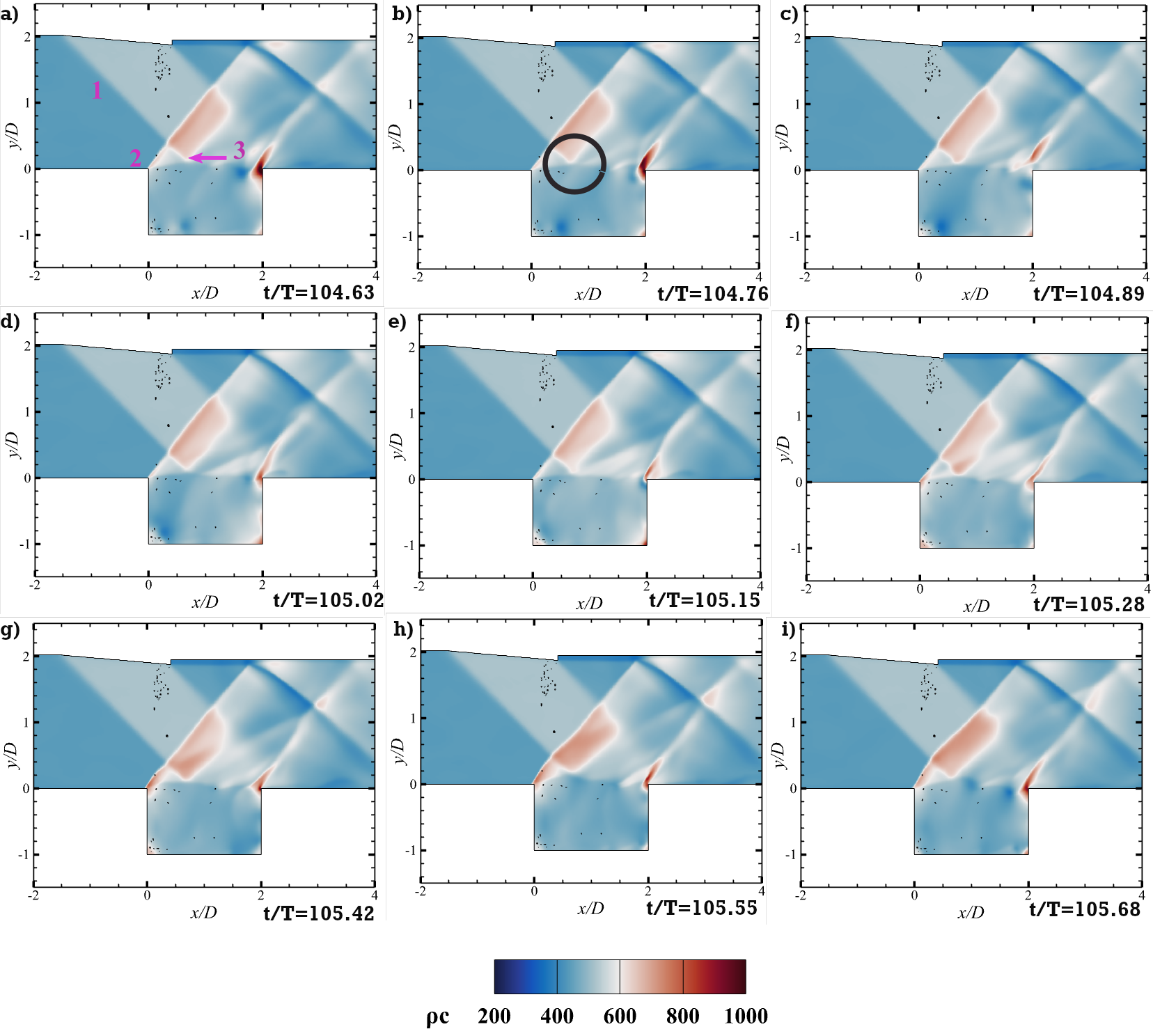}% Here is how to import EPS art
    \centering

    \caption{\label{fig:31}The acoustic impedance contour for the cavity configuration at $L/D=2$ at M$_\infty$=1.71. The feature marked as (1) represents the impinging shock, which first interacts with the leading-edge shock (2) before striking the shear layer. Upon interaction, the shear layer reflects the shock as an expansion wave (3). The region of decreasing impedance, highlighted by the black circle, facilitates the expansion process, supporting the flow divergence across (3). }
   \end{figure*}
   \begin{figure*}

	\includegraphics[scale=0.4]{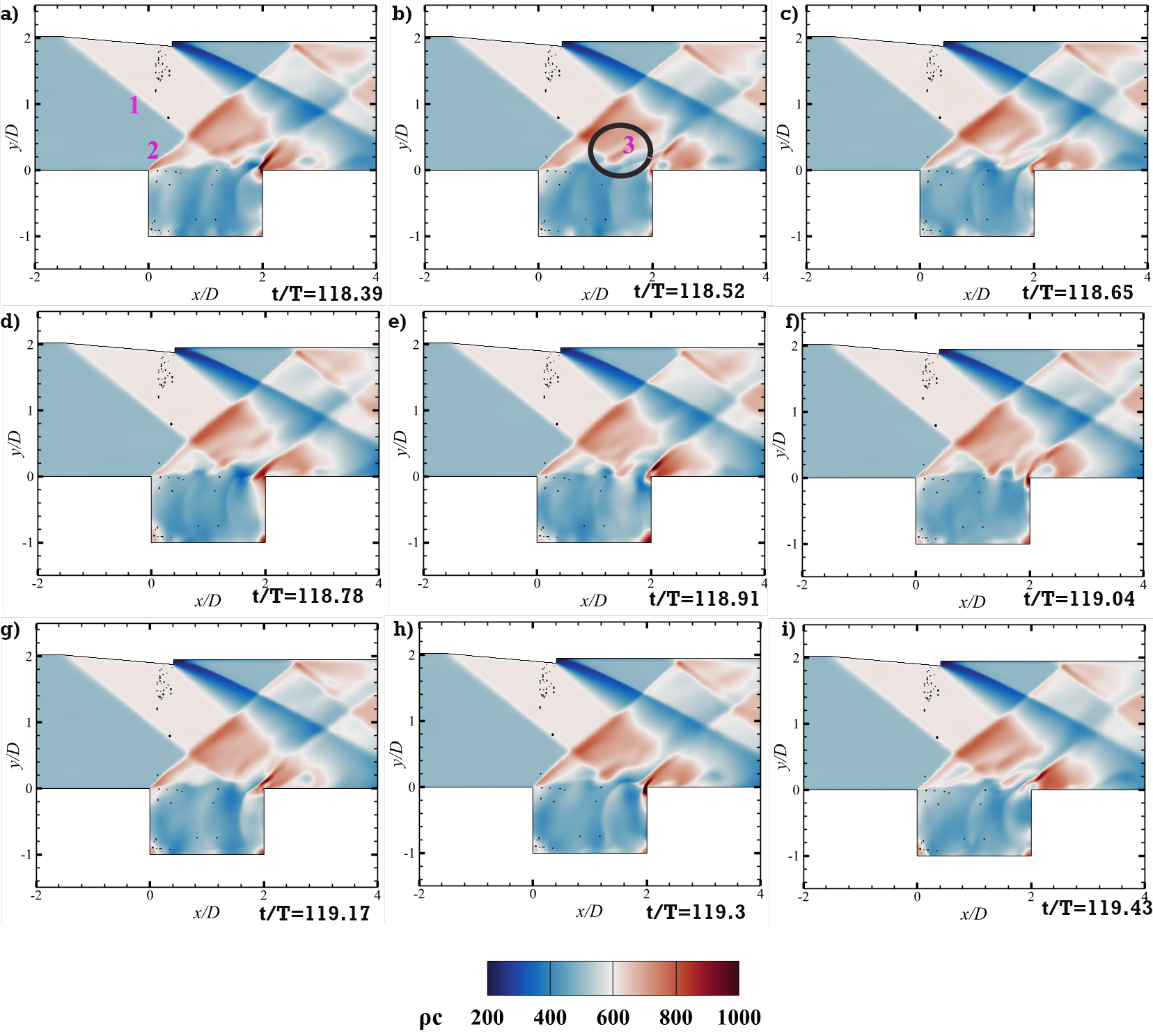}% Here is how to import EPS art
    \centering

    \caption{\label{fig:32}The acoustic impedance contour for the cavity configuration at $L/D=2$ at M$_\infty$=2. The feature marked as (1) represents the impinging shock, which first interacts with the leading-edge separation shock (2) before striking the shear layer. Upon interaction, the shear layer reflects the shock as an expansion wave (3). The region of decreasing impedance, highlighted by the black circle, facilitates the expansion process, supporting the flow divergence across (3). }
   \end{figure*}
   \begin{figure*}

	\includegraphics[scale=0.4]{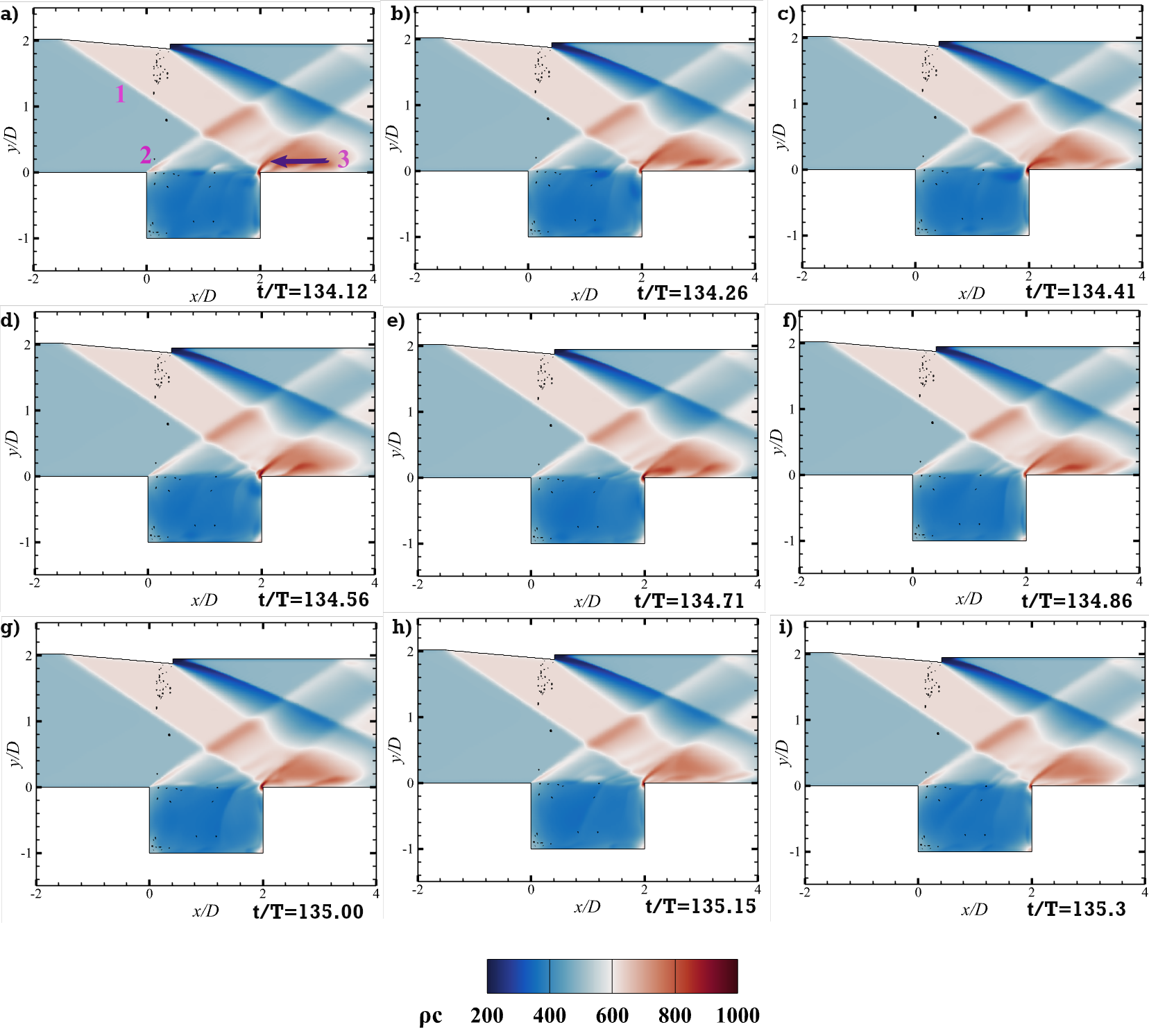}% Here is how to import EPS art
    \centering

    \caption{\label{fig:33} The acoustic impedance contour for the cavity configuration at $L/D=2$ at M$_\infty$=2.29. The feature marked as (1) represents the impinging shock, which first interacts with the leading-edge separation shock (2)  before striking the shear layer near the trailing edge of the cavity. A reattachment shock (3) forms near the aft wall as the shear layer impinges the aft wall.}
   \end{figure*}
   \begin{figure*}

	\includegraphics[scale=0.4]{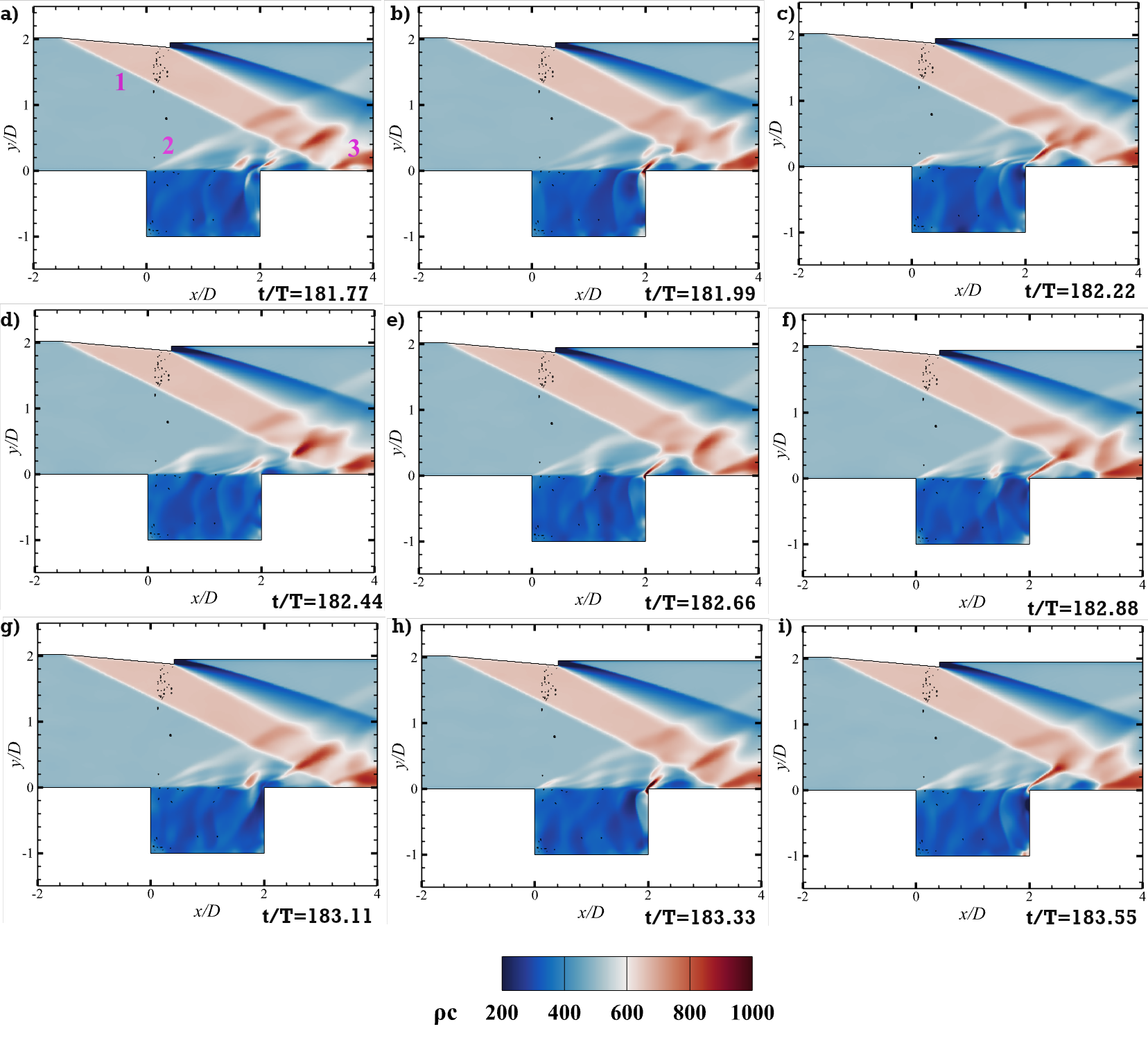}% Here is how to import EPS art
    \centering

    \caption{\label{fig:34}The acoustic impedance contour for the cavity configuration at $L/D=2$ at M$_\infty$=3. The feature marked as (1) represents the impinging shock, which first interacts with the leading-edge separation shock (2) before striking the surface downstream of the trailing edge. The impinging shock is reflected from the surface as another shock wave (3).  }
   \end{figure*}
Figures \ref{fig:31}–\ref{fig:34} illustrate the acoustic impedance for the cavity configuration at L/D=2 across all Mach numbers under study. For M$_\infty$ = 1.71 and 2 (Figs. \ref{fig:31}–\ref{fig:32}), the impinging shock reflects from the shear layer as an expansion fan, visualized as a sudden drop in impedance downstream of the impingement point.
At M$_\infty$ = 2.29 (Fig. \ref{fig:33}), a reattachment shock forms near the aft wall of the cavity. As the shock impinges on the shear layer close to the aft wall, its reflection as an expansion fan is overshadowed by the stronger reattachment shock, which dominates the local flow structure.
For M$_\infty$ = 3 (Fig. \ref{fig:34}), the impinging shock strikes the surface much farther downstream from the trailing edge of the cavity instead of the shear layer of the cavity. The surface then reflects it as another shock wave instead of an expansion fan.

\FloatBarrier  % Ensure floats are placed before moving to References

\section*{References}

\nocite{*}
\bibliography{aipsamp}% Produces the bibliography via BibTeX.

\end{document}